\pdfoutput=1
\documentclass[fleqn,useAMS,usenatbib]{mnras}
\usepackage{graphicx, siunitx, amssymb}
\usepackage[T1]{fontenc}
\usepackage{aecompl}

\DeclareSIUnit\gauss{G}
\DeclareSIUnit\year{yr}
\DeclareSIUnit\au{au}

\newcommand{\sups}[1]{$^{\rm #1}$}

\title[A BCool survey of the magnetic fields of planet-hosting solar-type stars]{A BCool survey of the magnetic fields of planet-hosting solar-type stars}
\author[M. W. Mengel et al.]{M. W. Mengel,$^{1}$\thanks{E-mail: matthew.mengel@usq.edu.au (MWM)} 
S. C. Marsden,$^{1}$ 
B. D. Carter,$^{1}$ 
J. Horner,$^{1}$
R. King,$^{1}$ \and
R. Fares,$^{2}$
S. V. Jeffers,$^{3}$
P. Petit,$^{4,5}$
A. A. Vidotto,$^{6}$
J. Morin,$^{7}$ \and
and the BCool Collaboration
\\
$^{1}$Computational Engineering and Science Research Centre, University of Southern Queensland, Toowoomba, Qld, Australia\\
$^{2}$INAF - Osservatorio Astrofisico di Catania, Via Santa Sofia, 78, I-95123 Catania, Italy \\
$^{3}$Institut f\"ur Astrophysik, Georg-August-Universit\"at G\"ottingen, Friedrich-Hund-Platz 1, 37077 G\"ottingen, Germany \\
$^{4}$Universit\'e de Toulouse, UPS-OMP, Institut de Recherche en Astrophysique et Plan\'etologie, F-31400 Toulouse, France \\ 
$^{5}$CNRS, Institut de Recherche en Astrophysique et Plan\'etologie, 14 Avenue Edouard Belin, F-31400 Toulouse, France \\
$^{6}$School of Physics, Trinity College Dublin, The University of Dublin, Dublin 2, Ireland \\
$^{7}$Laboratoire Univers et Particules de Montpellier, Universit\'e de Montpellier, CNRS, place Eug\'ene Bataillon, F-34095 Montpellier, France
}

\begin{document}

\date{Accepted 2016 November 11. Received 2016 November 10; in original form 2016 September 29}

\pagerange{\pageref{firstpage}--\pageref{lastpage}} \pubyear{2016}

\maketitle

\label{firstpage}


\begin{abstract}
We present a spectropolarimetric snapshot survey of solar-type planet hosting stars.  In addition to 14 planet-hosting stars observed as part of the BCool magnetic snapshot survey, we obtained magnetic observations of a further 19 planet-hosting solar-type stars in order to see if the presence of close-in planets had an effect on the measured surface magnetic field ($|B_{\ell}|$).  Our results indicate that the magnetic activity of this sample is congruent with that of the overall BCool sample.  The effects of the planetary systems on the magnetic activity of the parent star, if any, are too subtle to detect compared to the intrinsic dispersion and correlations with rotation, age and stellar activity proxies in our sample.  Four of the 19 newly observed stars, two of which are subgiants, have unambiguously detected magnetic fields and are future targets for Zeeman Doppler Mapping.
\end{abstract}

\begin{keywords}
line: profiles - stars: activity - stars: magnetic fields - techniques: polarimetric - planetary systems.
\end{keywords}

\setlength{\tabcolsep}{4pt}

\section{Introduction}

\subsection{The BCool Spectropolarimetric Survey}

The BCool spectropolarimetric survey \citep{Marsden2014} has the objective of observing a large sample of solar-type stars with $V \lesssim 9$, attempting to detect their magnetic fields.  This serves a two-fold purpose.  First, the survey sets out to determine if there is any correlation between the large-scale stellar magnetic field and various stellar parameters.  Secondly, the characterisation of the magnetic fields of the targeted stars allows for the selection of optimal or interesting targets for further study, such as long-term monitoring and mapping of their magnetic field topology in order to observe and characterise their magnetic cycles.

\citet{Marsden2014} published spectropolarimetric snapshots of 170 solar-type stars, and reported that the strength of the large-scale magnetic field declines as a function of age and with reduced rotation.  Additionally, they found that the mean surface magnetic field detected was higher for K-dwarfs compared to G-dwarfs and F-dwarfs.  \citet{Marsden2014} do note this higher field for K dwarfs may be due to selection biases, although this observation would also be consistent with even stronger fields seen on M dwarfs (e.g. \citealt{Morin2008}).  At the most fundamental level,  this work adds 19 additional solar-type stars to the BCool sample.

In this work, we also aim to examine whether the presence of a planetary system or the nature of the planets in that system correlates with the large-scale magnetic field of the host star.  Such correlation would potentially be indicative of star-planet interaction (SPI).

\subsection{Star-Planet Interaction}

Host stars profoundly influence their surrounding planetary environment.  The transfer of angular momentum, stellar winds and magnetic fields play roles in planetary formation, migration and evolution \citep{Horner2010}.  Additionally, stellar winds in cool stars, which are influenced by the variation in stellar magnetic fields, affect the planetary environment and can interact strongly with planetary atmospheres \citep{See2015,Strugarek2015,Vidotto2015}.  Understanding the magnetic field of host stars allows us to more fully investigate and understand the planetary environment around them, and by extension, examine the habitability of potential Earth-like planets.

Tidal and magnetic interaction between host stars and their planets, especially large close-in ``hot Jupiters'', has been of considerable interest since the discovery of such planets. Whether such companions result in any change in the magnetic activity of the star remains unanswered.  \citet{Cuntz2000}, \citet{Rubenstein2000} and \citet{Lanza2009,Lanza2012} variously suggest that close-in planets may spin up the host star by transfer of angular momentum resulting in higher activity.  Activity enhancement may also occur due to magnetic reconnection events between stellar and planetary magnetic fields.  Evidence for star-planet interaction has been claimed for several stars using a variety of observed phenomena synchronised with the orbital period of the planet, such as photospheric ``hot spots'' \citep{Lanza2011}, chromospheric enhancement \citep{Shkolnik2005,Shkolnik2008}, and X-ray enhancement \citep{Pillitteri2010}. 

A statistical assessment of SPI by \citet{Miller2015} shows that no particular correlation exists between proxies for SPI strength and coronal activity.  A relationship with solar type (FGK) stars was found, however they note that this is only driven by a handful of extreme hot Jupiter systems.  \citet{Miller2015} also investigated whether planetary properties were correlated with UV luminosity or Ca~\textsc{ii}~H\&K and found no significant difference between hot Jupiter systems and others.  This was in contrast to the conclusions drawn from earlier observations by \citet{Shkolnik2013} and \citet{Krejcova2012}.  However, all such studies note that selection effects may skew these results.  

Finally, \citet{France2016} presents tentative evidence for SPI for close-in, massive planets via an enhancement of the transition region.  \citet{France2016} speculate that this may be due to magnetospheric interaction, but urge caution due to a small sample size.

\section{Target Selection}

\subsection{New Targets (Not Previously Observed by BCool)}

Targets were chosen from the exoplanet.eu database\footnotemark\footnotetext{\url{http://exoplanet.eu}} \citep{Schneider2011}.  The host stars of the planetary systems were chosen to be broadly solar-type with $T_{\mbox{eff}}$ between \SI{5100}{\kelvin} and \SI{6300}{\kelvin} and $M_{\star} < 1.5 M_{\odot}$, on the main sequence or at the subgiant stage.  This selection was further constrained by the observational requirements of the NARVAL instrument and the T\'elescope Bernard Lyot (see Sec.~\ref{sec:obs:specpol}).  Targets were chosen with magnitude $V \lesssim 9$ and declination $\delta$ above \SI{-10}{\degree}. 

Table \ref{tab:planetParams} shows the configurations of the observed planetary systems.  There are six stars which host a planet that can be classified as a hot Jupiter (considering the definition of a massive planet with a semi-major axis of less than $\sim \SI{0.05}{\au}$).  Further, three systems contain two or more detected planets.

\begin{figure}
  \includegraphics[scale=0.45]{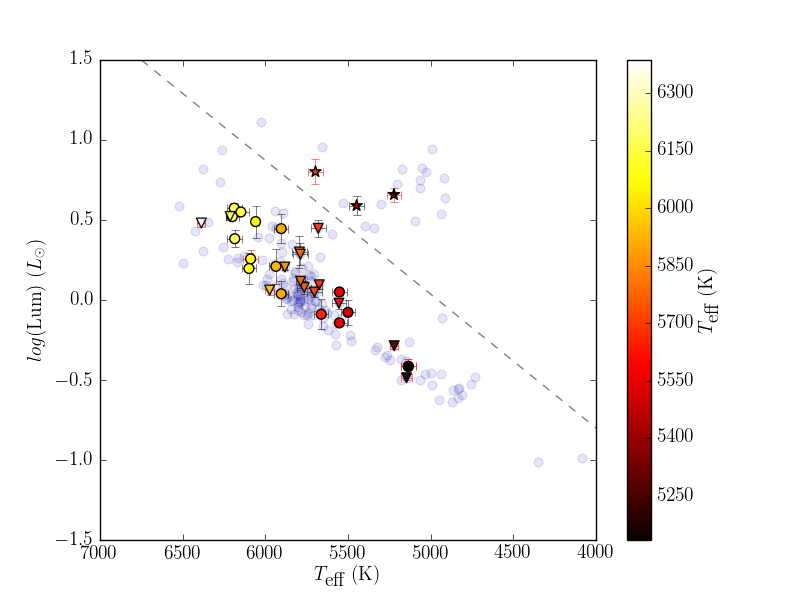}
  \includegraphics[scale=0.45]{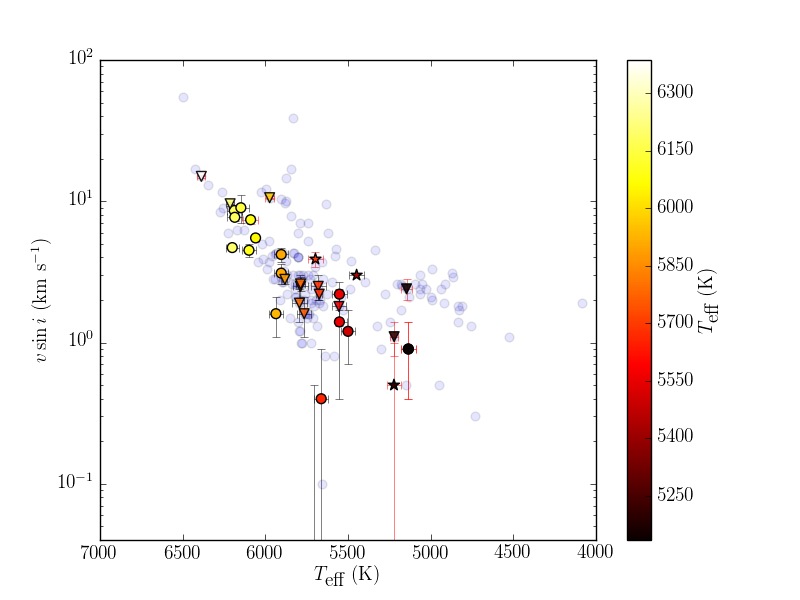}
  \caption{HR diagram (upper panel) and $v \sin i$ vs $T_{\mbox{eff}}$ plot (lower panel) for the survey stars (Table~\protect\ref{tab:stellarParams}; circular and star-shaped points) overlaid on the entire BCool sample (blue circles, data from \protect\citet[Fig. 1]{Marsden2014}). Planet-hosting stars from the BCool survey (Table~\protect\ref{tab:stellarParamsBC}) are shown as inverted triangles. Points with red error bars are magnetic detections, and the colour of the data points represents the effective temperature.  Five-pointed star-shaped data points indicate sub-giants and circles represent dwarfs with the dashed line dividing these two categories (note inverted triangles from the BCool survey are all dwarfs; no planet-hosts in the BCool sample were subgiants).  This dashed dividing line is in the same position as \protect\citet[Fig. 1]{Marsden2014}.}
  \label{fig:HR}
\end{figure}

The stellar parameters of the host stars are shown in Table~\ref{tab:stellarParams}.  12 of the 19 targets were included in the SPOCS \citep{Valenti2005} database, and thus their parameters were taken from \citet{Valenti2005} and \citet{Takeda2007}.  For the remaining targets, the stellar parameters were sourced from the references listed in the table.  Figure~\ref{fig:HR} shows the stars in our sample on a Hertzsprung-Russell (HR) diagram superimposed on the stars for the BCool sample \citep[Fig. 1]{Marsden2014}.  Figure~\ref{fig:HR} shows that the sample stars in this work are similar in general characteristics to the full BCool sample.  Three of the stars in our sample are classified as sub-giants with the remainder being dwarf stars.

The lower panel of Figure~\ref{fig:HR} shows the rotation velocity ($v \sin i$) for our sample against effective temperature, also superimposed on the overall BCool sample.  As shown, the sample exhibits a decrease in rotation rate with decreasing effective temperature.  Our entire sample of planet-hosting stars display $v \sin i < \SI{10}{\kilo\meter\per\second}$, which we note is at the lower end of the BCool sample.

\subsection{The BCool Sample}

In addition to the survey described above, we extracted the 14 known planet-hosting stars from the BCool sample.  As the methods used to derive their various parameters are the same as this new work, we did not re-analyse the data.  Instead, we use the results from \citet{Marsden2014} in our Discussion and Conclusions.  The planetary and stellar parameters are shown in Appendix~\ref{BCappendix}, as are the magnetic and chromospheric results from \citet{Marsden2014} for these stars.

\begin{table*}
\caption{Planetary parameters of the sample of planet-hosting solar type stars not previously observed by \protect\citet{Marsden2014}.  The stellar component's \textit{Hipparcos} number, SPOCS catalogue number and HD number (where applicable) are shown in the first three columns.  Column 4 refers to the Name by which the planetary components (column 5) are known.  The period, projected mass ($M \sin i$) and semi-major axis are shown for each planet.  These values are taken from the references listed in the last column; locations where values were unavailable in the literature are denoted by `$X$'.  Parameters for the systems observed by the BCool survey are shown in Table~\protect\ref{tab:planetParamsBC}.}
\label{tab:planetParams}
\begin{tabular}{lccccccccc}
\hline
\multicolumn{3}{c}{Star} &  & \multicolumn{6}{c}{Planet(s)} \\\cline{1-3}\cline{5-10}

\multicolumn{1}{c}{HIP} & \multicolumn{1}{c}{SPOCS} & \multicolumn{1}{c}{HD}  &  & \multicolumn{1}{c}{Name} & \multicolumn{1}{c}{Component} &  \multicolumn{1}{c}{Period} & \multicolumn{1}{c}{$M \sin i$} & \multicolumn{1}{c}{Semi-major} & \multicolumn{1}{c}{Refs.}  \\
\multicolumn{1}{c}{no.} & \multicolumn{1}{c}{no.}  & \multicolumn{1}{c}{no.} & &  \multicolumn{1}{c}{} & \multicolumn{1}{c}{} & \multicolumn{1}{c}{(\SI{}{\day})} & \multicolumn{1}{c}{($M_{J}$)} & \multicolumn{1}{c}{axis (\SI{}{\au})} & \multicolumn{1}{c}{}  \\
\hline \\[-1.5ex]
14954 & 155 & 19994 & & 94~Cet & b & $537.7 \pm 3.1$ & $1.69 \pm 0.26$ & $1.428 \pm 0.083$ & 1\\[3pt]
17747 & 184 & 23596 &  & HD~23596 & b & $1565 \pm 21$ & $7.8 \pm 1.1$ & $2.83 \pm 0.17$ & 1 \\[3pt]
24205$^{+}$ & 252 & 33636 & & HD~33636 & b & $2127.7 \pm 8.2$ & $9.28 \pm 0.77$ & $3.37 \pm 0.19$ & 1 \\[3pt]
25191 & ~ & 290327 &  & HD~290327 & b & $2443^{+205}_{-117}$ & $2.54^{+0.17}_{-0.14}$ & $3.43^{+0.20}_{-0.12}$ & 2  \\[3pt]
26381 & 270 & 37124 & & HD~37124 & b & $154.46 \pm X$ & $0.64 \pm 0.11$ & $0.529 \pm 0.031$ & 1 \\[3pt]
~ & ~ & ~ & & HD~37124 & c$^{*}$ & $2295.00 \pm X$ & $0.683 \pm 0.088$ & $3.19\pm0.18$ & 1 \\[3pt]
~ & ~ & ~ & & HD~37124 & d & $843.60 \pm X$ & $0.624 \pm 0.063$ & $1.639 \pm 0.095$ & 1 \\[3pt]
26664 & ~ & 37605 & & HD~37605 & b & $54.23 \pm 0.23$ & $2.86 \pm 0.41$ & $0.261 \pm 0.015$ & 1  \\[3pt]
27253 & 282 & 38529 & & HD~38529 & b & $14.3093 \pm 0.0013$ & $0.852 \pm 0.074$ & $0.1313 \pm 0.0076$ & 1 \\[3pt]
~ & ~ & ~ & & HD~38529 & c & $2165 \pm 14$ & $13.2 \pm 1.1$ & $3.74 \pm 0.22$ & 1 \\[3pt]
27384 & ~ &  38801 & & HD~38801 & b & $696.3 \pm 2.7$ & $10.7 \pm 0.5$ & $1.70 \pm 0.03$ & 3  \\[3pt]
28767 & 293 & 40979 & & HD~40979 & b & $263.84 \pm 0.71$ & $3.83 \pm 0.36$ & $0.855 \pm 0.049$ & 1 \\[3pt]
29301 & ~ & 42176 & & KELT2A & b & $4.11379 \pm 0.00001$ & $1.524 \pm 0.088$ & $0.05504 \pm 0.00086$ & 4 \\[3pt]
30057 & ~ & 43691 & & HD~43691 & b & $36.96 \pm 0.02$ & $2.49 \pm X$ & $0.24 \pm X$ & 5 \\[3pt]
32916 & 324 & 49674 & & HD~49674 & b & $4.94737 \pm 0.00098$ & $0.105 \pm 0.011$ & $0.0580 \pm 0.0034$ & 1 \\[3pt]
45406 & ~ & 79498 & & HD~79498 & b & $1966.1 \pm 41$ & $1.34 \pm 0.07$ & $3.13 \pm 0.08$ & 6 \\[3pt]
64457 & 556 & 114783 & & HD~114783 & b & $496.9 \pm 2.3$ & $1.034 \pm 0.089$ & $1.169 \pm 0.068$ & 1 \\[3pt]
95740 & 841 & 183263 & & HD~183263 & b & $635.4 \pm 3.9$ & $3.82 \pm 0.4$ & $1.525 \pm 0.088$ & 1 \\[3pt]
96507 & ~ & 185269 & & HD~185269 & b & $6.8399 \pm 0.0013$ & $1.03 \pm 0.03$ & $0.077 \pm X$ & 7 \\[3pt]
98767 & 870 & 190360 & & HD~190360 & b & $2891 \pm 85$ & $1.55 \pm 0.14$ & $3.99 \pm 0.25$ & 1 \\[3pt]
~ & ~ & ~ & ~ & HD~190360 & c & $17.100 \pm 0.015$ & $0.0587 \pm 0.0078$ & $0.1303 \pm 0.0075$ & 1 \\[3pt]

101966 & 901 & 196885 & & HD~196885 & b & $1326.0 \pm 3.7$ & $2.98 \pm 0.05$ & $2.6 \pm 0.1$ & 8 \\[3pt]

108859 & 953 & 209458 & & HD~209458 & b & $3.52474554 \pm 1.8\times10^{-7}$ & $0.689 \pm 0.057$ & $0.0474 \pm 0.0027$ & 1 \\[3pt]
\hline
\end{tabular} \\
References: 1: \cite{Butler2006}, 2: \cite{Naef2010}, 3: \cite{Harakawa2010}, 4: \cite{Beatty2012}, 5: \cite{DaSilva2007}, 6: \cite{Robertson2012}, 7: \cite{Moutou2006}, 8: \cite{Chauvin2011}. \\
$^{+}$ - HIP~24205 (HD~33636) at the time of writing is listed in some catalogues of planet-hosting stars, however \cite{Martioli2010} have determined the mass of the companion to be much too massive to be considered a planet.  While perhaps considered no longer planet-hosting, in the interests of completeness we include it in our analysis. \\
$^{*}$ -  \cite{Butler2006} indicates that the mass of HD~37124~c is unclear and an alternative interpretation is for a period of \SI{29.3}{\day}, $M \sin i = 0.170 M_{J}$ and semi-major axis of \SI{0.170}{\au}, with slightly different values for HD~37124~a and HD~37124~b.
\end{table*}

\begin{table*}
\scriptsize
\begin{center}
\caption{Stellar parameters of the sample of planet-hosting solar type stars not previously observed by \protect\citet{Marsden2014}.  The spectral type is taken from SIMBAD (\url{http://simbad.u-strasbg.fr/simbad/}, \protect\citet{Wenger2000}).  Column 11 is the radius of the convective zone of the star. Values are found in the references shown in the final column of the table; locations where values were unavailable in the literature are denoted by `$X$'.  $^{\textsc{SG}}$ indicates the star is a subgiant (see Fig.~\protect\ref{fig:HR}). Parameters for the systems observed by the BCool survey are shown in Table~\protect\ref{tab:stellarParamsBC}.}
\label{tab:stellarParams}
\begin{tabular}{lccccrccccccc}
\hline
\multicolumn{1}{c}{HIP} & \multicolumn{1}{c}{SPOCS} & \multicolumn{1}{c}{Spec.} & \multicolumn{1}{c}{$T_{\mbox{eff}}$} & \multicolumn{1}{c}{log($g$)} & \multicolumn{1}{c}{[M/H] or} & \multicolumn{1}{c}{log(Lum)} & \multicolumn{1}{c}{Age} & \multicolumn{1}{c}{Mass} & \multicolumn{1}{c}{Radius} & \multicolumn{1}{c}{Radius$_{\mbox{CZ}}$} & \multicolumn{1}{c}{$v \sin i$} & \multicolumn{1}{c}{Refs.} \\
\multicolumn{1}{c}{no.} & \multicolumn{1}{c}{no.}  & \multicolumn{1}{c}{Type} & \multicolumn{1}{c}{(\SI{}{\kelvin})} & \multicolumn{1}{c}{(\SI{}{\centi\meter\per\second\squared})} & \multicolumn{1}{c}{[Fe/H] ($^{*}$)} & \multicolumn{1}{c}{($L_{\sun}$)} & \multicolumn{1}{c}{(\SI{}{\giga\year})} & \multicolumn{1}{c}{($M_{\sun}$)} & \multicolumn{1}{c}{($R_{\sun}$)} & \multicolumn{1}{c}{($R_{\sun}$)} & \multicolumn{1}{c}{(\SI{}{\kilo\meter\per\second})} & \\
\hline \\[-1.5ex]
14954 & 155 & F8.5V & $6188^{+44}_{-44}$ & $4.17^{+0.03}_{-0.08}$ & $+0.17^{+0.03}_{-0.03}$ & $+0.574^{+0.035}_{-0.035}$ & $2.56^{+0.40}_{-0.36}$ & $1.365^{+0.042}_{-0.024}$ & $1.75^{+0.06}_{-0.16}$ & $0.265^{+0.009}_{-0.009}$ & $8.6^{+0.5}_{-0.5}$ & 1, 2 \\[3pt]
17747 & 184 & F8 & $5904^{+44}_{-44}$ & $4.07^{+0.04}_{-0.04}$ & $+0.24^{+0.03}_{-0.03}$ & $+0.446^{+0.089}_{-0.089}$ & $5.68^{+0.48}_{-0.36}$ & $1.159^{+0.062}_{-0.018}$ & $1.69^{+0.09}_{-0.08}$ & $0.472^{+0.035}_{-0.030}$ & $4.2^{+0.5}_{-0.5}$ & 1, 2\\[3pt]
24205 & 252 & G0 & $5904^{+44}_{-44}$ & $4.44^{+0.04}_{-0.04}$ & $-0.12^{+0.03}_{-0.03}$ & $+0.039^{+0.077}_{-0.077}$ & $3.52^{+2.16}_{-2.44}$ & $1.017^{+0.032}_{-0.032}$ & $1.02^{+0.04}_{-0.04}$ & $0.247^{+0.020}_{-0.015}$ & $3.1^{+0.5}_{-0.5}$ & 1, 2\\[3pt]
25191 & ~ & G0 & $5552^{+21}_{-44}$ & $4.42^{+0.04}_{-0.04}$ & $^{*}-0.11^{+0.02}_{-0.02}$ & $-0.143^{+X~~~}_{-X}$ & \multicolumn{1}{c}{$>3$} & $0.90~~^{+X~~~}_{-X~}$ & $1.00^{+0.01}_{-0.01}$ & \multicolumn{1}{c}{$X$} & $1.4^{+1.0}_{-1.0}$ & 10 \\[3pt]
26381 & 270 & G4IV-V & $5500^{+44}_{-44}$ & $4.44^{+0.04}_{-0.02}$ & $-0.29^{+0.03}_{-0.03}$ & $-0.077^{+0.077}_{-0.077}$ & $11.7^{+3.1~~}_{-8.4}$ & $0.850^{+0.022}_{-0.016}$ & $0.93^{+0.03}_{-0.04}$ & $0.277^{+0.011}_{-0.018}$ & $1.2^{+0.5}_{-0.5}$ & 1, 2 \\[3pt]
26664$^{\textsc{sg}}$ & ~ & K0 &  $5448^{+44}_{-44}$ & $4.51^{+0.02}_{-0.02}$ &  $^{*}+0.34^{+0.03}_{-0.03}$  & $+0.590^{+0.058}_{-0.058}$ & $7.07^{+X~~}_{-X}$ & $1.000^{+0.050}_{-0.050}$ & $0.90^{+0.05}_{-0.05}$ & \multicolumn{1}{c}{$X$} & \multicolumn{1}{c}{$<1$} & 3, 4 \\[3pt]
27253$^{\textsc{sg}}$ & 282 & G8III/IV & $5697^{+44}_{-44}$ & $3.94^{+0.02}_{-0.02}$ & $+0.27^{+0.03}_{-0.03}$ & $+0.802^{+0.079}_{-0.079}$ & $3.28^{+0.36}_{-0.24}$ & $1.477^{+0.040}_{-0.052}$ & $2.50^{+0.08}_{-0.06}$ & $0.711^{+0.031}_{-0.014}$ & $3.9^{+0.5}_{-0.5}$ & 1, 2\\[3pt]
27384$^{\textsc{sg}}$ & ~ & G8IV & $5222^{+44}_{-44}$ & $3.84^{+0.10}_{-0.10}$ & $^{*}+0.26^{+0.03}_{-0.03}$ & $+0.659^{+0.043}_{-0.047}$ & $4.67^{+2.56}_{-2.56}$ & $1.36~~^{+0.09~}_{-0.09~}$ & $2.53^{+0.13}_{-0.13}$ & \multicolumn{1}{c}{$X$} & $0.5^{+0.5}_{-0.5}$ & 5 \\[3pt]
28767 & 293 & F8 & $6089^{+44}_{-44}$ & $4.32^{+0.04}_{-0.03}$ & $+0.12^{+0.03}_{-0.03}$ & $+0.257^{+0.055}_{-0.055}$ & $3.56^{+0.68}_{-0.80}$ & $1.154^{+0.028}_{-0.022}$ & $1.23^{+0.05}_{-0.04}$ & $0.273^{+0.020}_{-0.018}$ & $7.4^{+0.5}_{-0.5}$ & 1, 2 \\[3pt]
29301 & ~ & F7V & $6148^{+48}_{-48}$ & $4.03^{+0.02}_{-0.03}$ & $^{*}+0.03^{+0.08}_{-0.08}$ & $+0.550^{+X~~~}_{-X}$ & $3.97^{+0.01}_{-0.01}$ & $1.314^{+0.063}_{-0.060}$ & $1.84^{+0.07}_{-0.05}$ & \multicolumn{1}{c}{$X$} & $9.0^{+2.0}_{-2.0}$ & 11, 13 \\[3pt]
30057 & ~ & G0 & $6200^{+40}_{-40}$ & $4.28^{+0.13}_{-0.13}$ & $^{*}+0.28^{+0.05}_{-0.05}$ & $+0.521^{+X~~~}_{-X}$ & $2.8~~^{+0.8~}_{-0.8}$ & $1.38~~^{+0.05~}_{-0.05~}$ & \multicolumn{1}{c}{$X$} & \multicolumn{1}{c}{$X$} & $4.7^{+X~}_{-X~}$ & 6, 13 \\[3pt]
32916 & 324 & G0 & $5662^{+44}_{-44}$ & $4.51^{+0.03}_{-0.03}$ & $+0.22^{+0.03}_{-0.03}$ & $-0.089^{+0.094}_{-0.094}$ & \multicolumn{1}{c}{$X$} & $1.015^{+0.048}_{-0.036}$ & $0.95^{+0.04}_{-0.04}$ & $0.271^{+0.013}_{-0.009}$ & $0.4^{+0.5}_{-0.5}$ & 1, 2 \\[3pt]
45406 & ~ & G5 & $5760^{+80}_{-80}$ & $4.37^{+0.12}_{-0.12}$ & $^{*}+0.24^{+0.06}_{-0.06}$ & \multicolumn{1}{c}{$X$} & $2.70^{+X~~}_{-X}$ & $1.06~~^{+X~~~}_{-X~}$ & \multicolumn{1}{c}{$X$} & \multicolumn{1}{c}{$X$} & \multicolumn{1}{c}{$X$} & 7, 8 \\[3pt]
64457 & 556 & K1V & $5135^{+44}_{-44}$ & $4.57^{+0.03}_{-0.04}$ & $+0.10^{+0.03}_{-0.03}$ & $-0.415^{+0.045}_{-0.045}$ & $6.76^{+X~~}_{-X}$ & $0.853^{+0.034}_{-0.038}$ & $0.81^{+0.02}_{-0.03}$ & $0.255^{+0.020}_{-0.010}$ & $0.9^{+0.5}_{-0.5}$ & 1, 2 \\[3pt]
95740 & 841 & G2IV & $5936^{+44}_{-44}$ & $4.36^{+0.05}_{-0.05}$ & $+0.22^{+0.03}_{-0.03}$ & $+0.210^{+0.110}_{-0.110}$ & $4.52^{+0.76}_{-1.12}$ & $1.121^{+0.064}_{-0.040}$ & $1.18^{+0.07}_{-0.07}$ & $0.304^{+0.026}_{-0.023}$ & $1.6^{+0.5}_{-0.5}$ & 1, 2 \\[3pt]
96507 & ~ & G2V & $6059^{+18}_{-18}$ & $4.13^{+0.06}_{-0.06}$ & $^{*}+0.12^{+0.02}_{-0.02}$ & $+0.49~^{+0.1~}_{-0.1}$ & $3.4~~^{+0.54}_{-0.54}$ & $1.33~~^{+0.07~}_{-0.07}$ & $1.76^{+0.07}_{-0.07}$ & \multicolumn{1}{c}{$X$} & $5.5^{+X~}_{-X~}$ & 9, 12 \\[3pt]
98767 & 870 & G7IV-V & $5552^{+44}_{-44}$ & $4.31^{+0.03}_{-0.02}$ & $+0.19^{+0.03}_{-0.03}$ & $+0.050^{+0.022}_{-0.022}$ & $13.4^{+X}_{-1.84}$ & $0.983^{+0.026}_{-0.048}$ & $1.15^{+0.03}_{-0.03}$ & $0.397^{+0.008}_{-0.022}$ & $2.2^{+0.5}_{-0.5}$ & 1, 2 \\[3pt]
101966 & 901 & F8IV & $6185^{+44}_{-44}$ & $4.26^{+0.03}_{-0.03}$ & $+0.13^{+0.03}_{-0.03}$ & $+0.383^{+0.054}_{-0.054}$ & $3.12^{+0.36}_{-0.4}$ & $1.230^{+0.028}_{-0.020}$ & $1.38^{+0.06}_{-0.05}$ & $0.270^{+0.020}_{-0.019}$ & $7.7^{+0.5}_{-0.5}$ & 1, 2 \\[3pt]
108859 & 953 & G0V & $6099^{+44}_{-44}$ & $4.39^{+0.04}_{-0.04}$ & $+0.02^{+0.03}_{-0.03}$ & $+0.197^{+0.095}_{-0.095}$ & $2.44^{+1.32}_{-1.64}$ & $1.131^{+0.026}_{-0.024}$ & $1.14^{+0.05}_{-0.05}$ & $0.241^{+0.018}_{-0.016}$ & $4.5^{+0.5}_{-0.5}$ & 1, 2 \\[3pt]
\hline
\end{tabular}
\end{center}
References: 1: \cite{Valenti2005}, 2: \cite{Takeda2007}, 3: \cite{Wang2012}, 4: \cite{Isaacson2010}, 5: \cite{Harakawa2010}, 6: \cite{DaSilva2007}, 7: \cite{Casagrande2011}, 8: \cite{Robertson2012}, 9: \cite{Moutou2006}, 10: \cite{Naef2010}, 11: \cite{Beatty2012}, 12: \cite{Jofre2015}, 13: \cite{McDonald2012}. \\
\end{table*}

\section{Observations and Data Processing}

\subsection{Instrument and Observational Procedure}
\label{sec:obs:specpol}

Observations were obtained using the polarimetric mode of the NARVAL spectropolarimeter (attached to the T\'elescope Bernard Lyot located at Observatoire du Pic du Midi).  NARVAL is composed of a bench mounted high resolution spectrograph and a Cassegrain-mounted polarimetry module.  The spectrograph has an optical wavelength coverage of \SIrange[range-units = single]{370}{1000}{\nano\meter}, and a resolution of $\sim65000$ with a pixel size of \SI{2.6}{\kilo\meter\per\second}.

The polarimetric module performs polarimetry over the entire spectral range using a series of three Fresnel rhombs.  The light is then split into two beams containing opposite polarisation states.  These two beams are fed via individual fibres to the spectrograph, allowing the simultaneous capture of both polarisation states and further allowing the unpolarised Stokes $I$ and circularly-polarised Stokes $V$ spectra to be determined from each observation.  Further information on NARVAL can be found in \citet{Auriere2003}.

Each Stokes $V$ observation consists of a sequence of four individual exposures.  Effectively, this results in eight individual spectra; four left-hand and four right-hand circularly polarised.  As described by \citet{Petit2003}, the polarisation states in the fibre pair described above are alternated during the sequence to help eliminate instrumental effects; the first and fourth exposures have one arrangement of polarisation states whilst the second and third have the opposite arrangement.  Adding all eight spectra yields the unpolarised Stokes $I$ (intensity) spectrum.  The polarised Stokes $V$ spectrum is obtained as per \citet[Eqs. 1, 2]{Donati1997b}:

\begin{equation}
 {V \over I} = {{R_{V} - 1} \over {R_{V} + 1}}
  \label{eq:voveri}
\end{equation}
where

\begin{equation}
 {R_{V}^{4}} = {{i_{1,\perp}/{i_{1,\parallel}} \over {i_{2,\perp}}/{i_{2,\parallel}}}~{i_{4,\perp}/{i_{4,\parallel}} \over {i_{3,\perp}}/{i_{3,\parallel}}}}
  \label{eq:r4}
\end{equation}
and $i_{k,\perp}$ and $i_{k,\parallel}$ are the two polarised spectra in each exposure, $k$.

By destructively adding the spectra, a null polarisation spectrum, $N$, can be obtained \citep[Eqs. 1, 3]{Donati1997b}:

\begin{equation}
 {N \over I} = {{R_{N} - 1} \over {R_{N} + 1}}
  \label{eq:noveri}
\end{equation}
where

\begin{equation}
 {R_{N}^{4}} = {{i_{1,\perp}/{i_{1,\parallel}} \over {i_{4,\perp}}/{i_{4,\parallel}}}~{i_{2,\perp}/{i_{2,\parallel}} \over {i_{3,\perp}}/{i_{3,\parallel}}}}
  \label{eq:r4N}
\end{equation}

As described in \citet{Bagnulo2009}, a significant signal (i.e. deviation from zero) in the $N$ spectrum may be indicative of a spurious polarisation signal.

\subsection{Spectropolarimetric Observations}

All observations were made using NARVAL at T\'elescope Bernard Lyot between September 2014 and January 2015.  Each star was observed once, except for HIP~64457, which was observed twice.  Each observation consisted of a spectropolarimetric sequence of four \SI{900}{\second} exposures.  The journal of observations is shown in Table~\ref{tab:obsJournal}.

\begin{table}
\caption{Journal of Observations showing the object, date and time of the observation, and the exposure time used.  As explained in Sec.~\protect\ref{sec:obs:specpol}, a Stokes $V$ observation consists of a sequence of four exposures, hence the nomenclature used here.}
\label{tab:obsJournal}
\begin{tabular}{lcccc}
\hline
\multicolumn{1}{c}{HIP} & \multicolumn{1}{c}{HJD} & \multicolumn{1}{c}{Date} & \multicolumn{1}{c}{\textsc{ut}} & \multicolumn{1}{c}{$T_{\mbox{exp}}$} \\
\multicolumn{1}{c}{no.} & \multicolumn{1}{c}{2450000$+$} & \multicolumn{1}{c}{}  & \multicolumn{1}{c}{hh:mm:ss} & \multicolumn{1}{c}{(\SI{}{\second})} \\
\hline \\[-1.5ex]
14954 & 7033.33090 & 2015-01-10 & 19:53:06 & $4 \times 900$ \\
17747 & 6958.50071 & 2014-10-27 & 23:53:43 & $4 \times 900$ \\
24205 & 6926.67026 & 2014-09-26 & 04:02:50 & $4 \times 900$ \\
25191 & 7034.43181 & 2015-01-11 & 22:15:20 & $4 \times 900$ \\
26381 & 6957.55553 & 2014-10-27 & 01:14:29 & $4 \times 900$ \\
26664 & 6962.64066 & 2014-11-01 & 03:17:03 & $4 \times 900$ \\
27253 & 6982.58476 & 2014-11-21 & 01:55:21 & $4 \times 900$ \\
27384 & 6960.64135 & 2014-10-30 & 03:19:08 & $4 \times 900$ \\
28767 & 6927.68776 & 2014-09-27 & 04:27:50 & $4 \times 900$ \\
29301 & 6994.63288 & 2014-12-03 & 03:04:50 & $4 \times 900$ \\
30057 & 6994.71807 & 2014-12-03 & 05:08:10 & $4 \times 900$ \\
32916 & 6959.64396 & 2014-10-29 & 03:23:27 & $4 \times 900$ \\
45406 & 6995.69091 & 2014-12-04 & 04:31:58 & $4 \times 900$ \\
64457 & 7030.70798 & 2015-01-08 & 04:59:21 & $4 \times 900$ \\
 & 7032.69851 & 2015-01-10 & 04:45:26 & $4 \times 900$ \\
95740 & 6995.28708 & 2014-12-03 & 18:58:48 & $4 \times 900$ \\
96507 & 6982.30247 & 2014-11-20 & 19:18:19 & $4 \times 900$ \\
98767 & 6982.35222 & 2014-11-20 & 20:29:11 & $4 \times 900$ \\
101966 & 6984.27534 & 2014-11-22 & 18:38:31 & $4 \times 900$ \\
108859 & 6957.33207 & 2014-10-26 & 19:53:29 & $4 \times 900$ \\
\hline
\end{tabular}
\end{table}

\subsection{Data Processing}

Observations were automatically reduced by a pipeline process utilising the \textsc{libre-esprit} package.  \textsc{libre-esprit} is based on the \textsc{esprit} software \citep{Donati1997b}.  The reduced Stokes $I$ and Stokes $V$ spectra were produced using \SI{1.8}{\kilo\meter\per\second} pixel resolution.

\subsection{Least Squares Deconvolution}

Zeeman signatures are typically very small and usually the S/N in a reduced spectrum is insufficient for a detection in a single line \citep{Donati1992}.  Least Squares Deconvolution (LSD) is a multiline technique which extracts Stokes $I$ and Stokes $V$ information from each individual spectral line in the reduced spectrum and determines an average profile with a higher S/N than for each individual line \citep{Donati1992,Kochukhov2010}.

\citet{Marsden2014} computed a set of line masks for use in LSD for the BCool sample using the Vienna Atomic Line Database (VALD\footnotemark\footnotetext{\url{http://vald.astro.unive.ac.at/\~vald/php/vald.php}}; \citealp{Kupka2000}).  These masks are derived from stellar atmospheric and spectral synthesis models using the stellar parameters $T_{eff}$, $\log(g)$ and log(M/H) (or $\log$(Fe/H) if $\log$(M/H) is unavailable or unknown).  For consistency, these same masks were used in this work, and the range of parameters and step size used is shown in Table~\ref{tab:maskParams}. For more details on the creation of the set of line masks, see \citet[Sec.~3.3]{Marsden2014}.

\begin{table}
\caption{Stellar paramters used to generate line masks for use in LSD. From \protect\citet[Table~2]{Marsden2014}.}
\label{tab:maskParams}
\begin{tabular}{lccc}
\hline
\multicolumn{1}{l}{Parameter} & Units & \multicolumn{1}{c}{Range} & \multicolumn{1}{c}{Step size} \\
\hline \\[-1.5ex]
$T_{eff}$ & \SI{}{\kelvin} & $4000-6500$ & 250 \\
$\log(g)$ & \SI{}{\centi\meter\per\second\squared} & $3.5-4.5$ & 0.5 \\
log(M/H) & & $-0.2~\mbox{to}~+0.2$ & 0.2 \\
\hline
\end{tabular}
\end{table}

For each target, the appropriate line mask was chosen, and LSD profiles for Stokes $I$ and Stokes $V$ were created with a resolved element of \SI{1.8}{\kilo\meter\per\second}.  Depending on the stellar parameters of the star in our sample, the number of lines used in the LSD process varies from $\sim7000$ to $\sim14000$.

As in \citet{Marsden2014}, the weighting of the spectral lines was adjusted such that the mean weights of the Stokes $V$ and Stokes $I$ profiles were close to unity.  We use the equations given by \citet[Eqs.~3, 4, 5]{Marsden2014} for calculating mean weights, and the same normalisation parameters ($d_{0}$ = line central depth, $\lambda_{0}$ = line central wavelength, $g_{0}$ = line Land\'e factor), varied for each \SI{250}{\kelvin} step in effective temperature.  The normalisation parameters are additionally used in the calculation of the longitudinal magnetic field (Sec.~\ref{sec:longMag}) and are reproduced in Table~\ref{tab:normParams}. 

\begin{table}
\caption{Normalization parameters used to produce LSD profiles, following \protect\citet[Table~4]{Marsden2014}.  $d_{0}$ = line central depth, $\lambda_{0}$ = line central wavelength, $g_{0}$ = line Land\'e factor}
\label{tab:normParams}
\begin{tabular}{lccc}
\hline
\multicolumn{1}{l}{$T_{eff}(\SI{}{\kelvin})$} & $d_{0}$ & \multicolumn{1}{c}{$\lambda_{0}$(\SI{}{\nano\meter})} & \multicolumn{1}{c}{$g_{0}$} \\
\hline \\[-1.5ex]
4000 & 0.55 & 650.0 & 1.22 \\
4250 & 0.55 & 640.0 & 1.22 \\
4500 & 0.55 & 630.0 & 1.22 \\
4750 & 0.55 & 620.0 & 1.22 \\
5000 & 0.54 & 610.0 & 1.22 \\
5250 & 0.54 & 600.0 & 1.22 \\
5500 & 0.53 & 590.0 & 1.22 \\
5750 & 0.52 & 580.0 & 1.22 \\
6000 & 0.51 & 570.0 & 1.22 \\
6250 & 0.50 & 570.0 & 1.21 \\
6500 & 0.49 & 560.0 & 1.21 \\
\hline
\end{tabular}
\end{table}

\subsection{The Longitudinal Magnetic Field}
\label{sec:longMag}

The mean longitudinal magnetic field, $B_{\ell}$ (or given as $\langle B_{\mbox{z}} \rangle$ in some publications) is the line-of-sight component of the stellar magnetic field integrated over the visible disc of the star.  $B_{\ell}$ can be obtained from the Stokes $I$ and Stokes $V$ LSD profiles.  From \citet{Donati1997b, Mathys1989}, for the given velocity ($v$, in \SI{}{\kilo\meter\per\second}) space:

\begin{equation}
B_{\ell} = -2.14 \times 10^{11} {{ \int v  V(v) dv} \over { \lambda_{0} g_{0} c \int [I_c - I(v)] dv }}
  \label{eq:Bl}
\end{equation}
where $B_{\ell}$ is in gauss, and $\lambda_{0}$ and $g_{0}$ are given in Table~\ref{tab:normParams}.  $c$ is the speed of light in \SI{}{\kilo\meter\per\second} and $I_{c}$ is the continuum level of the Stokes $I$ LSD profile and is normalised to 1.  The error in $B_{\ell}$ ($B_{err}$) is calculated by propagating the uncertainties in the reduction pipeline through equation~\ref{eq:Bl}.  As mentioned by \citet{Marsden2014} and discussed in depth by \citet[Sec.~5]{Shorlin2002}, the uncertainty depends upon the S/N of the observation, the number of lines used to produce the Stokes $V$ profile, and the depth and width of the average line.  The line depth and width scale linearly with $v~\sin~i$.

An additional uncertainty in $B_{\ell}$ is introduced by the choice of the velocity domain used to integrate equation~\ref{eq:Bl}.  A narrow velocity domain potentially excludes polarised signals, while a domain which is too wide potentially introduces spurious signals due to the noise in the Stokes $V$ spectrum.  For consistency with the measurement of $B_{\ell}$ of the BCool sample, the method outlined by \citet{Marsden2014} was used. $B_{\ell}$ was calculated using a range of velocity domains, following which, the domain for which the ratio of $|B_{\ell}|/B_{err}$ was maximised was chosen.

For each observation, the value of $B_{\ell}$ was calculated for the null profile ($N_{\ell}$) over the same velocity domain used for the Stokes $V$ profile.  A value of $|N_{\ell}|$ which is close to zero is indicative that the magnetic field measurement is unaffected by spurious polarisation signals.  This is indeed the case for the majority of our sample.  Where $|N_{\ell}|$ departs significantly from zero, and is large relative to $|B_{\ell}|$, the measurement must be considered carefully.  The values of $B_{\ell}$ and $N_{\ell}$ for each observation, and the velocity domain used for the calculations are presented in Table~\ref{tab:magneticField}.

\subsection{Magnetic Detection}
\label{sec:FAP}

\citet{Donati1992} and \citet{Donati1997b} describe a method of determining whether a magnetic field is ``detected'' on a star using the Stokes $V$ profile.  That is, a probability is calculated as to whether the variations in the Stokes $V$ LSD profile are likely due to the presence of a magnetic field rather than from noise.

Reduced $\chi^{2}$ statistics are calculated for the Stokes $V$ and $N$ profiles, inside and outside the spectral lines as defined by the position of the unpolarised Stokes $I$ profile in velocity space.  From these values a False Alarm Probability (FAP) is determined.  An unambiguous (or definite) detection is defined as having $\mbox{FAP} < 10^{-5}$ (corresponding to a $\chi^{2}$ probability greater than 99.999 percent).   A marginal detection is defined as having $10^{-5} < \mbox{FAP} < 10^{-3}$ ($\chi^{2}$ probability between 99.999 and 99.9 percent).  It is to be noted that there are no marginal detections in our sample.  $\mbox{FAP} > 10^{-3}$ is classified as a non-detection.

For each observation, the FAP and the classification as a definite detection (D) or non-detection (N) is shown in Table~\ref{tab:magneticField}.  It should be noted that irrespective of the detection state, values for $B_{\ell}$  and $N_{\ell}$ are shown.  The probability function used to determine the FAP takes into account the Stokes $V$ and $N$ both inside and outside the spectral lines, whereas the values of $N_{\ell}$ and $B_{\ell}$ are calculated only over a velocity domain containing the spectral lines.  Thus detections depend on the false alarm probability rather than absolute longitudinal field values.  In general, for non-detections (N), $B_{\ell}$ will be of the same order as $N_{\ell}$, i.e the probability is that the measured $B_{\ell}$ is due to noise in the Stokes $V$ LSD profile rather than magnetic activity.  The exception in our sample is HIP~28767 (see Table~\ref{tab:magneticField} and Appendix~\ref{sec:HIP28767}), which has a value of $N_{\ell}$ close to the calculated $B_{\ell}$, but a very low FAP (i.e. similar values of $N_{\ell}$ and $B_{\ell}$ do not necessarily preclude a detection).  

\subsection{Radial Velocities}

\begin{figure}
  \includegraphics[scale=0.45]{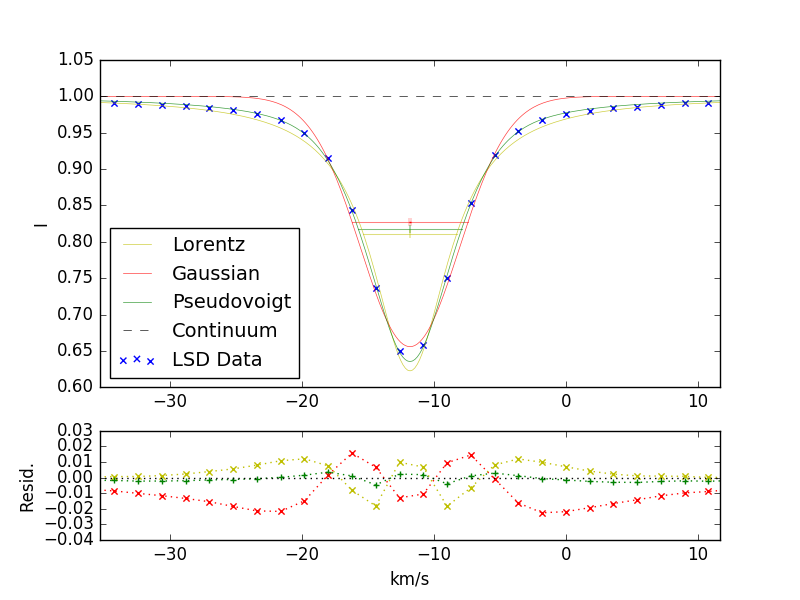}
  \caption{Determining the radial velocity of HIP~64457 (HD~114783).  The data from the unpolarised Stokes $I$ LSD profile is shown by the blue crosses.  Gaussian (red), Lorentz (yellow) and Pseudovoigt (green) profiles are fitted to the LSD profile data (upper panel).  Residuals from each fit are shown in the lower panel.  The centroid of the Pseudovoigt fitted profile is taken as the radial velocity of the star.}
  \label{fig:rvHIP64457}
\end{figure}

The radial velocities of the host stars of the planetary systems were determined in this work (Table~\ref{tab:magneticField}) and compared, where available, to those calculated by \citet{Nidever2002}.  A Pseudovoigt profile was fitted to the Stokes $I$ LSD profile of each observation, with the centroid of the Pseudovoigt considered to be the radial velocity (e.g. Fig~\ref{fig:rvHIP64457}).  The long-term radial velocity stability of NARVAL is \SI{30}{\meter\per\second} \citep{Moutou2007}.

While our values are generally close to those previously published, it must be noted that since our sample stars host planetary systems, there will be intrinsic variation in the radial velocity of the host star due to the gravitational influence of the planets.  Thus differences, potentially of significant magnitude, are to be expected depending upon the specific time of observation and the orbital configurations.  Hence the values are presented as indicative only, and for completeness with the BCool sample.

\subsection{Stellar Activity Proxies}

As for the BCool sample, various measures of stellar activity were calculated for each target in our sample.  In additon to the Ca~\textsc{ii}~H\&K S-index \citep{Wright2004}, Ca~\textsc{ii} infrared triplet (IRT) and H$\alpha$ indices were derived.

\subsubsection{Ca \textsc{ii} H \& K Emission (S-index)}

Following the methodology of \citet{Wright2004}, \citet{Marsden2014} determined for NARVAL the coefficients $a$, $b$, $c$, $d$, and $e$ (see Table~\ref{tab:scoeffs}) of the equation: 

\begin{equation}
 \mbox{S-index} = {{aF_H + bF_K} \over {cF_{R_{HK}} + dF_{V_{HK}} }} + e
  \label{eq:sindex}
\end{equation}
where $F_H$ and $F_K$ are the fluxes in \SI{2.18}{\angstrom} triangular bandpasses centred on the cores of the Ca~\textsc{ii}~H\&K lines and $F_{R_{HK}}$ and $F_{V_{HK}}$ are the fluxes in two rectangular \SI{20}{\angstrom} bandpasses centred on the continuum either side of the HK lines at \SI{3901.07}{\angstrom} and \SI{4001.07}{\angstrom} respectively.

For each reduced unpolarised spectrum of each star, overlapping orders were removed.  Equation~\ref{eq:sindex} was then applied to the remaining spectrum, generating the Ca~\textsc{ii}~H\&K S-indices for each spectrum.  As per \citet{Marsden2014}, the sample standard deviation of the S-indices for each star was calculated as an empirical measure of the uncertainty.  The mean and standard deviation values for each star are shown in Table~\ref{tab:derivedValues}.  

\subsubsection{Derived Chromospheric Parameters}

The S-indices were then used to derive various chromospheric parameters for each star, all of which are shown in Table~\ref{tab:derivedValues}.  $\log(R^{\prime}_{HK})$ was derived using \citet[eqs. 9-12]{Wright2004}, using \textit{Hipparcos} $B-V$ values.  Using the further formulations from \citet[eqs. 13-15]{Wright2004}, $\log({P_{rot}/\tau})$ (log(Rossby number)), the chromospheric period and chromospheric age for each target were calculated.

\subsubsection{H$\alpha$ Emission}

The H$\alpha$-index was determined for each unpolarised reduced spectrum using the equation:

\begin{equation}
 \mbox{H}\alpha{\mbox{-index}} = {{F_{H\alpha}} \over {F_{V_{H\alpha}} + F_{R_{H\alpha}} }}
  \label{eq:haindex}
\end{equation}
where $F_{H\alpha}$ is the flux from a \SI{3.6}{\angstrom} rectangular bandpass centred on the H$\alpha$ line (\SI{6562.85}{\angstrom}), and $F_{V_{H\alpha}}$ and $F_{R_{H\alpha}}$ are the fluxes in \SI{2.2}{\angstrom} rectangular bandpasses located on the continuum either side of the H$\alpha$ line centred at 6558.85 and \SI{6567.30}{\angstrom}.  These bandpasses are defined by \citet[Table~3]{Gizis2002}.

As for the S-index, the sample standard deviation of the H$\alpha$-indices for each star was calculated as an empirical measure of the uncertainty.  The mean and standard deviation of the H$\alpha$-index for each star are shown in Table~\ref{tab:derivedValues}.

\subsubsection{Ca \textsc{ii} IRT Emission}

The activity index for the Ca~\textsc{ii}~infrared triplet ($\mbox{Ca}_{\textsc{IRT}}$-index) was calculated from each unpolarised reduced spectrum using the equation \citep[eq.~1]{Petit2013}:

\begin{equation}
 \mbox{Ca}_{\textsc{IRT}}{\mbox{-index}} = {{F_{8498} + F_{8542} + F_{8662}} \over {F_{V_{IRT}} + F_{R_{IRT}} }}
  \label{eq:cairt}
\end{equation}
where $F_{8498}$, $F_{8542}$ and $F_{8662}$ are the integrated fluxes of three \SI{2}{\angstrom} rectangular bandpasses centred on the corresponding Ca~\textsc{ii}~IRT lines (located at 8498.02, 8542.09 and \SI{8662.14}{\angstrom}), while $F_{V_{IRT}}$ and $F_{R_{IRT}}$ are the fluxes in \SI{5}{\angstrom} rectangular bandpasses located on the continuum either side of the Ca~\textsc{ii} triplet at 8475.8 and \SI{8704.9}{\angstrom}.  The mean and standard deviation of the $\mbox{Ca}_{\textsc{IRT}}$-index for each star are shown in Table~\ref{tab:derivedValues}.

\section{Results and Discussion}

\begin{table*}
\scriptsize
\caption{Results from the analysis of the Stokes $V$ LSD profiles of the stars not observed by \protect\citet{Marsden2014}.  Column 3 provides the date of the observation (in the case of HIP~64457 two observations were obtained).  Column 4 shows the radial velocity for the star (see \protect\ref{fig:rvHIP64457}); for comparison column 5 shows the radial velocity measured by \protect\citet{Nidever2002} or in two cases from \protect\citet{Valenti2005} indicated by $^{\textsc{VF}}$.  Columns 6 and 7 show the signal-to-noise of the Stokes $V$ profile and the number of lines used in the LSD process respectively.  Column 8 indicates is the magnetic field was unambiguously detected (D) or not (N); note we have no marginal detections, represented by M in \protect\citet[Table~3]{Marsden2014}.  Column 9 shows the false alarm probability calculated for the detection in column 8.  Columns 10 and 11 indicate the velocity range used to calculate $B_{\ell}$ (column 12) and $N_{\ell}$ (column 13) using equation \protect\ref{eq:Bl}.}
\label{tab:magneticField}
\begin{tabular}{lcccccccccccccc}
\hline
\multicolumn{1}{c}{HIP} & \multicolumn{1}{c}{Obs.} & \multicolumn{1}{c}{HJD} & \multicolumn{1}{c}{RV} & \multicolumn{1}{c}{RV} & \multicolumn{1}{c}{SNR$_{\mbox{LSD}}$} & \multicolumn{1}{c}{lines} & \multicolumn{1}{c}{Detection} & \multicolumn{1}{c}{FAP} & \multicolumn{2}{c}{Velocity} & \multicolumn{1}{c}{$B_{\ell}$} & \multicolumn{1}{c}{$N_{\ell}$}  \\
\multicolumn{1}{c}{no.} & \multicolumn{1}{c}{no.} & \multicolumn{1}{c}{+2450000}  & \multicolumn{1}{c}{(this work)} & \multicolumn{1}{c}{(Nidever)} & \multicolumn{1}{c}{} & \multicolumn{1}{c}{used} & \multicolumn{1}{c}{} & \multicolumn{1}{c}{} & \multicolumn{2}{c}{range (\SI{}{\kilo\meter\per\second})} & \multicolumn{1}{c}{(\SI{}{\gauss})} & \multicolumn{1}{c}{(\SI{}{\gauss})}  \\
\multicolumn{1}{c}{} & \multicolumn{1}{c}{} & \multicolumn{1}{c}{}  & \multicolumn{1}{c}{(\SI{}{\kilo\meter\per\second})} & \multicolumn{1}{c}{(\SI{}{\kilo\meter\per\second})} & \multicolumn{1}{c}{} & \multicolumn{1}{c}{} & \multicolumn{1}{c}{} & \multicolumn{1}{c}{} & \multicolumn{2}{c}{} & \multicolumn{1}{c}{} & \multicolumn{1}{c}{}  \\
\hline \\[-1.5ex]
14954 & 1 & 7033.33090 & $+19.64$ & $+19.331$ & 60048 & 11101 & N & $5.320 \times 10^{-1~}$ & $+9$ & $+31$ & $+0.4 \pm 0.3$ & $~~0.0 \pm 0.3$ \\[3pt]
17747 & 1 & 6958.50071 & $-9.97$ & $-10.6^{\textsc{VF}}$ & 18483  & 10602 & N & $9.967 \times 10^{-1~}$ & $-25$ & $+5$ & $+1.2 \pm 1.1$ & $+1.5 \pm 1.1$ \\[3pt]
24205 & 1 & 6926.67026 & $+5.74$ & $+5.714$ & 16164 & 7584 & N & $2.109 \times 10^{-1~}$ & $+0$ & $+13$ & $+1.3 \pm 0.4$ & $+0.3 \pm 0.4$ \\[3pt]
25191 & 1 & 7034.43181 & $+29.54$ & $-$ & 6046 & 9392 & N & $4.518 \times 10^{-1~}$ & $+18$ & $+41$ & $-2.3 \pm 1.9$ & $+1.9 \pm 1.9$ \\[3pt]
26381 & 1 & 6957.55553 & $-22.93$  & $-23.076$ & 12834 & 9365 & N & $6.022 \times 10^{-1~}$ & $-34$ & $-13$ & $-1.3 \pm 1.2$ & $+2.3 \pm 1.2$ \\[3pt]
26664 & 1 & 6962.64066 & $-21.85$ & $-$ & 11787 & 12097 & N & $1.245 \times 10^{-1~}$ & $-27$ & $-16$ & $-1.5 \pm 0.5$ & $+0.3 \pm 0.5$ \\[3pt]
27253 & 1 & 6982.58476 & $+30.44$ & $+30.210$ & 28583 & 12803 & D & $2.509 \times 10^{-9~}$ & $+23$ & $+38$ & $+1.7 \pm 0.3$ & $+0.0 \pm 0.3$ \\[3pt]
27384 & 1 & 6960.64135 & $-25.29$ & $-$ & 13887 & 13918 & D & $1.332 \times 10^{-15}$ & $-31$ & $-20$ & $+3.4 \pm 0.4$ & $-0.1 \pm 0.4$ \\[3pt]
28767 & 1 & 6927.68776 & $+32.85$ & $+32.542$ & 20086 & 10062 & D & $1.448 \times 10^{-9~}$ & $+14$ & $+52$ & $+4.5 \pm 1.4$ & $+3.5 \pm 1.4$ \\[3pt]
29301 & 1 & 6994.63288 & $-47.25$ & $-$ & 8003 & 8309 & N & $4.801 \times 10^{-1~}$ & $-59$ & $-36$ & $-2.7 \pm 1.7$ & $+1.0 \pm 1.7$ \\[3pt]
30057 & 1 & 6994.71807 & $-28.98$ & $-$ & 12649 & 9147 & N & $9.741 \times 10^{-1~}$ & $-36$ & $-22$ & $+0.5 \pm 0.6$ & $-0.5 \pm 0.6$ \\[3pt]
32916 & 1 & 6959.64396 & $+12.18$ & $+12.045$ & 14420 & 11084 & N & $3.266 \times 10^{-2~}$ & $+5$ & $+18$ & $+0.9 \pm 0.5$ & $-0.9 \pm 0.5$ \\[3pt]
45406 & 1 & 6995.69091 & $+20.08$ & $-$ & 14833 & 11058 & N & $9.316 \times 10^{-1~}$ & $+9$ & $+32$ & $-1.5 \pm 1.0$ & $+2.1 \pm 1.0$ \\[3pt]
64457 & 1 & 7030.70798 & $-11.80$ & $-12.012$ & 17873 & 13110 & D & $6.762 \times 10^{-7~}$ & $-18$ & $-5$ & $+2.5 \pm 0.4$ & $-0.5 \pm 0.4$ \\[3pt]
           & 2 & 7032.69851 & $-11.75$ & $-12.012$ & 21789 & 13048 & D & $5.749 \times 10^{-13}$ & $-18$ & $-5$ & $+2.4 \pm 0.3$ & $-0.6 \pm 0.3$ \\[3pt]
95740 & 1 & 6995.28708 & $-50.16$ & $-54.9^{\textsc{VF}}$ & 9290 & 10006 & N & $1.839 \times 10^{-1~}$ & $-58$ & $-41$ & $+3.0 \pm 0.9$ & $+0.6 \pm 0.9$ \\[3pt]
96507 & 1 & 6982.30247 & $+0.82$ & $-$ & 14747 & 10618 & N & $5.742 \times 10^{-1~}$ & $-7$ & $+9$ & $+0.7 \pm 0.7$ & $+0.9 \pm 0.7$ \\[3pt]
98767 & 1 & 6982.35222 & $-45.08$ & $-45.308$ & 25620 & 12051 & N & $9.512 \times 10^{-1~}$ & $-50$ & $-40$ & $-0.2 \pm 0.2$ & $+0.3 \pm 0.2$ \\[3pt]
101966 & 1 & 6984.27534 & $-30.80$ & $-30.189$ & 14961 & 9143 & N & $9.940 \times 10^{-1~}$ & $-45$ & $-16$ & $+1.7 \pm 1.23$ & $-0.3 \pm 1.3$ \\[3pt]
108859 & 1 & 6957.33207 & $-14.56$ & $-14.759$ & 11851 & 7164 & N & $4.666 \times 10^{-1~}$ & $-23$ & $-5$ & $-0.9 \pm 0.8$ & $-0.6 \pm 0.8$ \\[3pt]
\hline
\end{tabular}
\end{table*}

\begin{table}
 \centering
  \caption{Table of coefficients for Equation~\protect\ref{eq:sindex} as calculated by \protect\cite{Marsden2014} for the NARVAL instrument.}
\label{tab:scoeffs}
  \begin{tabular}{@{}ccc@{}}
  \hline
   Coefficient & NARVAL \\
 \hline
$a$ & 12.873 \\
$b$ & 2.502 \\
$c$ & 8.877 \\
$d$ & 4.271 \\
$e$ & $1.183 \times 10^{-3}$ \\
\hline
\end{tabular}
\end{table}

\begin{table*}
\scriptsize
\caption{Chromospheric activity of the stars not observed by \protect\citet{Marsden2014}.  $B-V$ and $V$ values are from \textit{Hipparcos}.  Where \protect\citet{Wright2004} has calculated an S-index, this is shown in column 3.  Chromospheric ages, periods and $\log(P_{rot}/\tau)$ have been determined using the equations presented in \protect\citet{Wright2004}.  As noted in the text, sample standard deviations are used as an indication in of various errors.  Note that HIP~64457 ($^{+}$) has two sequences of observations (8 exposures) compared with all other targets (4 exposures).}
\label{tab:derivedValues}
\begin{tabular}{lcccccccccccccc}
\hline
\multicolumn{1}{c}{HIP} & \multicolumn{1}{c}{$B-V$} & \multicolumn{1}{c}{$V$} & \multicolumn{1}{c}{S-index} & \multicolumn{1}{c}{S-index} & \multicolumn{1}{c}{log($R^{\prime}_{HK}$)} & \multicolumn{1}{c}{Chromospheric} & \multicolumn{1}{c}{Chromospheric} & \multicolumn{1}{c}{Ca$_{\mbox{IRT}}$-index} & \multicolumn{1}{c}{H${_{\alpha}}$-index} & \multicolumn{1}{c}{$\log(P_{rot}/\tau)$} \\
\multicolumn{1}{c}{no.} & \multicolumn{2}{c}{(Hipparcos)} & \multicolumn{1}{c}{Wright.}  & \multicolumn{1}{c}{(this work)} & \multicolumn{1}{c}{} & \multicolumn{1}{c}{Age (\SI{}{\giga\year})} & \multicolumn{1}{c}{period (\SI{}{\day})} & \multicolumn{1}{c}{} & \multicolumn{1}{c}{} & \multicolumn{1}{c}{} & \multicolumn{1}{c}{} \\
\hline \\[-1.5ex]

14954 & 0.575 & 5.07 & 0.173 & $0.1569 \pm 0.0003$ & $-4.99^{+0.00}_{-0.00}$ & $5.404^{+0.051}_{-0.044}$ & $16.1^{+0.0}_{-0.0}$ & $0.7411 \pm 0.0011$ & $0.3057 \pm 0.0001$ & $+0.320^{-0.001}_{+0.001}$ \\[3pt]
17747 & 0.634 & 7.25 & 0.150 & $0.1451 \pm 0.0004$ & $-5.10^{+0.00}_{-0.01}$ & $7.814^{+0.132}_{-0.079}$ & $25.4^{+0.1}_{-0.1}$ & $0.7561 \pm 0.0008$ & $0.3106 \pm 0.0001$ & $+0.361^{-0.002}_{+0.001}$ \\[3pt]
24205 & 0.588 & 7.00 & 0.180 & $0.1773 \pm 0.0011$ & $-4.87^{+0.01}_{-0.00}$ & $3.515^{+0.049}_{-0.105}$ & $15.5^{+0.1}_{-0.2}$ & $0.8209 \pm 0.0009$ & $0.3146 \pm 0.0002$ & $+0.266^{-0.002}_{+0.004}$ \\[3pt]
25191 & 0.761 & 8.99 & $-$ & $0.1558 \pm 0.0085$ & $-5.05^{+0.05}_{-0.07}$ & $6.777^{+1.622}_{-1.069}$ & $39.4^{+2.4}_{-1.6}$ & $0.8139 \pm 0.0031$ & $0.3307 \pm 0.0019$ & $+0.344^{-0.026}_{+0.018}$ \\[3pt]
26381 & 0.667 & 7.68 & 0.179 & $0.1881 \pm 0.0016$ & $-4.86^{+0.01}_{-0.01}$ & $3.314^{+0.078}_{-0.078}$ & $23.5^{+0.2}_{-0.2}$ & $0.8462 \pm 0.0007$ & $0.3292 \pm 0.0006$ & $+0.257^{-0.004}_{+0.004}$ \\[3pt]
26664 & 0.827 & 8.67 & $-$ & $0.1866 \pm 0.0049$ & $-4.94^{+0.02}_{-0.02}$ & $4.524^{+0.397}_{-0.253}$ & $40.3^{+1.0}_{-0.6}$ & $0.7546 \pm 0.0004$ & $0.3401 \pm 0.0008$ & $+0.299^{-0.010}_{+0.007}$ \\[3pt]
27253 & 0.773 & 5.95 & 0.174 & $0.1750 \pm 0.0005$ & $-4.96^{+0.00}_{-0.00}$ & $4.854^{+0.056}_{-0.036}$ & $37.3^{+0.1}_{-0.1}$ & $0.6871 \pm 0.0007$ & $0.3231 \pm 0.0001$ & $+0.308^{-0.001}_{+0.001}$ \\[3pt]
27384 & 0.873 & 8.26 & $-$ & $0.1978 \pm 0.0046$ & $-4.93^{+0.01}_{-0.02}$ & $4.393^{+0.290}_{-0.240}$ & $42.3^{+0.8}_{-0.7}$ & $0.7187 \pm 0.0006$ & $0.3368 \pm 0.0005$ & $+0.296^{-0.008}_{+0.007}$ \\[3pt]
28767 & 0.573 & 6.74 & 0.234 & $0.2543 \pm 0.0003$ & $-4.58^{+0.00}_{-0.00}$ & $1.249^{+0.002}_{-0.005}$ & $7.8^{+0.0}_{-0.0}$ & $0.8398 \pm 0.0017$ & $0.3138 \pm 0.0002$ & $+0.011^{-0.001}_{+0.002}$ \\[3pt]
29301 & 0.530 & 8.68 & $-$ & $0.1384 \pm 0.0028$ & $-5.13^{+0.04}_{-0.01}$ & $8.752^{+0.392}_{-1.088}$ & $12.9^{+0.2}_{-0.5}$ & $0.7734 \pm 0.0010$ & $0.3041 \pm 0.0008$ & $+0.375^{-0.006}_{+0.016}$ \\[3pt]
30057 & 0.596 & 8.03 & $-$ & $0.1492 \pm 0.0019$ & $-5.05^{+0.01}_{-0.02}$ & $6.763^{+0.466}_{-0.290}$ & $19.6^{+0.3}_{-0.2}$ & $0.7451 \pm 0.0006$ & $0.3076 \pm 0.0002$ & $+0.345^{-0.008}_{+0.005}$ \\[3pt]
32916 & 0.729 & 8.10 & 0.211 & $0.1858 \pm 0.0012$ & $-4.90^{+0.01}_{-0.01}$ & $3.824^{+0.090}_{-0.076}$ & $31.0^{+0.2}_{-0.2}$ & $0.7615 \pm 0.0009$ & $0.3281 \pm 0.0003$ & $+0.278^{-0.003}_{+0.003}$ \\[3pt]
45406 & 0.693 & 8.05 & $-$ & $0.1493 \pm 0.0013$ & $-5.08^{+0.01}_{-0.01}$ & $7.346^{+0.286}_{-0.262}$ & $32.6^{+0.3}_{-0.3}$ & $0.7730 \pm 0.0009$ & $0.3222 \pm 0.0003$ & $+0.354^{-0.004}_{+0.004}$ \\[3pt]
64457$^{+}$ & 0.930 & 7.56 & 0.215 & $0.2033 \pm 0.0032$ & $-4.96^{+0.01}_{-0.01}$ & $4.896^{+0.207}_{-0.184}$ &$45.4^{+0.5}_{-0.5}$ & $0.7618 \pm 0.0018$ & $0.3528 \pm 0.0007$ & $+0.309^{-0.005}_{+0.004}$ \\[3pt]
95740 & 0.678 & 7.86 & 0.145 & $0.1521 \pm 0.0027$ & $-5.06^{+0.02}_{-0.02}$ & $6.811^{+0.450}_{-0.509}$ & $30.1^{+0.5}_{-0.6}$ & $0.7634 \pm 0.0007$ & $0.3171 \pm 0.0005$ & $+0.346^{-0.007}_{+0.008}$ \\[3pt]
96507 & 0.606 & 6.67 & $-$ & $0.1286 \pm 0.0009$ & $-5.26^{+0.01}_{-0.02}$ & $12.638^{+0.559}_{-0.214}$ & $25.4^{+0.5}_{-0.2}$ & $0.7486 \pm 0.0021$ & $0.3110 \pm 0.0001$ & $+0.430^{-0.008}_{+0.003}$ \\[3pt]
98767 & 0.749 & 5.73 & 0.148 & $0.1499 \pm 0.0018$ & $-5.08^{+0.01}_{-0.01}$ & $7.461^{+0.354}_{-0.321}$ & $39.2^{+0.5}_{-0.4}$ & $0.7657 \pm 0.0019$ & $0.3324 \pm 0.0003$ & $+0.356^{-0.005}_{+0.005}$ \\[3pt]
101966 & 0.559 & 6.39 & 0.151 & $0.1455 \pm 0.0014$ & $-5.07^{+0.01}_{-0.01}$ & $7.213^{+0.332}_{-0.334}$ & $15.5^{+0.2}_{-0.2}$ & $0.7701 \pm 0.0010$ & $0.3039 \pm 0.0006$ & $+0.352^{-0.005}_{+0.005}$ \\[3pt]
108859 & 0.594 & 7.65 & 0.154 & $0.1558 \pm 0.0085$ & $-5.01^{+0.06}_{-0.09}$ & $5.831^{+1.886}_{-1.193}$ & $18.6^{+1.5}_{-1.0}$ & $0.8025 \pm 0.0023$ & $0.3112 \pm 0.0003$ & $+0.326^{-0.033}_{+0.024}$ \\[3pt]

\hline

\end{tabular}
\end{table*}

\subsection{Magnetic Detections}

Due to the small size of our sample, we did not attempt a comprehensive analysis of detection rates and their correlation with stellar and observational parameters as per \citep{Marsden2014}.  However, we can make some broad observations relating to the overall BCool sample.

Of the sample of 19 newly-observed stars, we obtained magnetic detections with a $\mbox{FAP} < 10^{-5}$ on four targets.  As noted by \citet{Marsden2014} for the BCool sample in general, the detection rates of the magnetic field drop with increasing age, with decreasing $v \sin i$, and with decreasing activity (i.e. S-index).  Given that our sample of planet-hosting stars are generally older than \SI{2}{\giga\year}, are relatively slow rotators, \textit{and} have low activity indices, it is to be expected that our overall detection rate will be low.  As such, it is unsurprising that it is in fact lower than the overall BCool sample.  Using the broad categorization of \citet{Marsden2014} (G-stars: $\SI{5000}{\kelvin} \leq T_{eff} \leq \SI{6000}{\kelvin}$; F-Stars: $T_{eff} > \SI{6000}{\kelvin}$) we have a 25 percent detection rate for G-stars (cf. 38\% BCool) and 14 percent for F-stars (cf. 32\% BCool).

Higher measured $B_{\ell}$ correlates with a higher S-index, and magnetic detections are more prevalent with higher S-indices ($\gtrsim 0.18$).  However, some magnetic detections with lower S-indices are apparent; these exceptions generally have a significantly higher SN$_{\mbox{LSD}}$.  This correlation is also noted by \citet{Marsden2014}.  

Even if we consider the additional planet-hosting stars from the BCool survey (Appendix~\ref{BCappendix}), our detection rate remains very low (24\% of the 33 total stars).  Our rate is even lower if the two very young planet-hosts HIP~16537 ($\epsilon$~Eri; $0.00^{+0.60}_{-0.00} \SI{}{\giga\year}$) and HIP~107350 (HN~Peg; $0.00^{+0.88}_{-0.00} \SI{}{\giga\year}$) from the BCool survey are disregarded as outliers, given that high activity (and thus magnetic field) is strongly correlated with young age and rapid rotation.

We find a correlation with S-index.  All stars with an S-index greater than $\sim0.18$ obtained a detection.   Only one star of our 19 new targets, HIP~27253, with an S-index below 0.18 ($0.1750 \pm 0.0005$) produced a detection.  This result is consistent with the finding from \citet{Marsden2014} that, as would be expected, increasing S-index is strongly correlated with the rate of detections.

\subsection{$|B_{\ell}|$ Measurements}

\begin{figure}
  \includegraphics[scale=0.45]{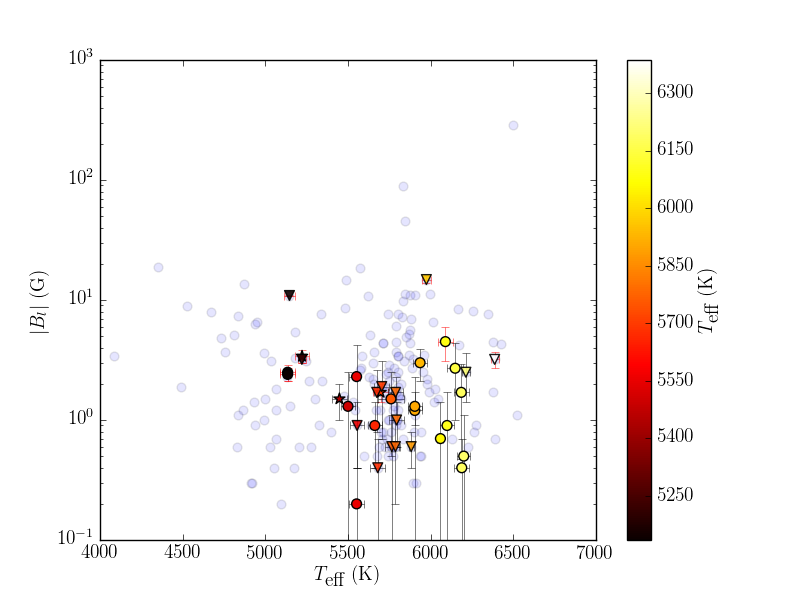}
  \caption{Plot of the maximum measured $|B_{\ell}|$ vs $T_{\mbox{eff}}$ for the planet-hosting stars (symbols are as in Fig.~\protect\ref{fig:HR}).  The complete BCool sample is shown as blue circles.}
  \label{fig:BvsT}
\end{figure}

In Figure~\ref{fig:BvsT}, we plot the maximum $|B_{\ell}|$ against $T_{eff}$ for the sample of 19 newly-observed stars and the BCool planet-hosts superimposed on the entire BCool sample (data from \citet{Marsden2014}).  Our results are consistent with the low activity area of the Bcool sample.

\begin{figure}
  \includegraphics[scale=0.45]{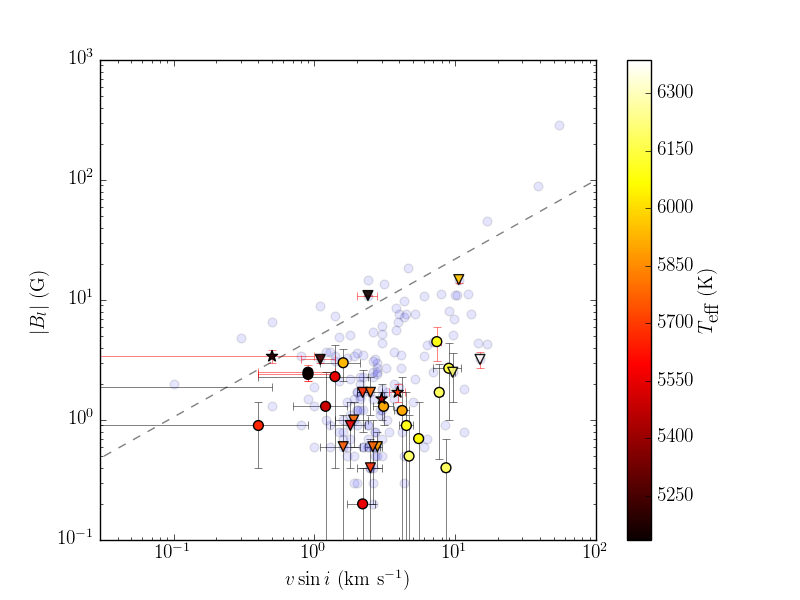}
  \caption{Plot of the maximum measured $|B_{\ell}|$ vs $v \sin i$ for the planet-hosting stars (symbols are as in Fig.~\protect\ref{fig:HR}). Stars located above the dashed line have significantly higher $|B_{\ell}|$ than others with similar $v \sin i$, as per \protect\citet[Sec.~6.3.3]{Marsden2014}. The complete BCool sample is shown as blue circles.}
  \label{fig:BvsVSINI}
\end{figure}

In Figure~\ref{fig:BvsVSINI}, we similarly overplot our sample and BCool planet-hosts over the entire BCool survey for $|B_{\ell}|$ against $v \sin i$.  As in \citet[Fig.~13, Sec~6.3.3]{Marsden2014}, stars above the dashed line are considered to have high values of $|B_{\ell}|$ compared to stars with a similar $v \sin i$.  We note that only a single star in our new sample (HIP~28767) has a longitudinal magnetic field marginally above this line.  It also has, by far, the highest S-index in our sample.  Given the correlation of decreasing $|B_{\ell}|$ with increasing age it is not surprising that almost our entire sample is located below this cutoff line.

\begin{figure}
  \includegraphics[scale=0.45]{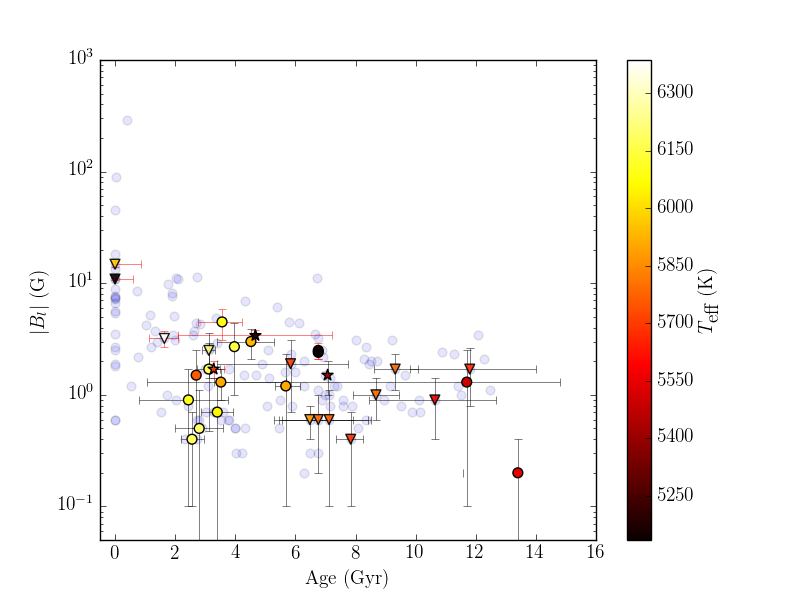}
  \includegraphics[scale=0.45]{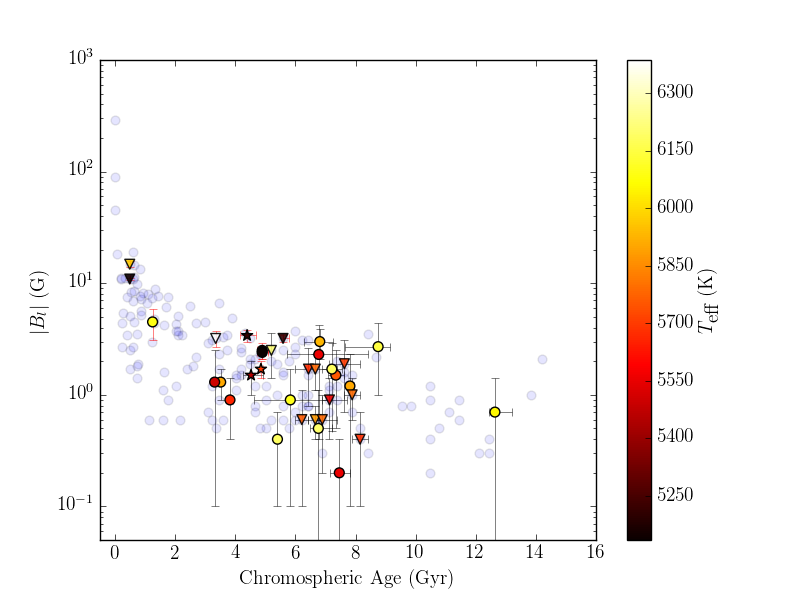}
  \caption{Plot of maximum measured $|B_{\ell}|$ vs Age (upper panel) and Chromospheric Age (lower panel; calculated from the equations of \protect\citet{Wright2004} and shown in Table~\protect\ref{tab:derivedValues}) for the planet-hosting stars. Symbols are as in Fig.~\protect\ref{fig:HR}. The complete BCool sample is shown as blue circles.}
  \label{fig:Ages}
\end{figure}

Figure~\ref{fig:Ages} illustrates that once again, the planet hosting sample of this work is entirely consistent with the BCool survey results.  In the upper panel of Figure~\ref{fig:Ages}, we show the magnetic field strength against the published ages of the stars (superimposed on the BCool sample) and in the lower panel, the magnetic field against the chromospheric age we derive. It is clear that our sample has older, evolved stars, and as expected their level of activity is lower than the younger stars which generate the more extreme magnetic fields.

Another general observation consistent with \citet{Marsden2014} is that cooler stars in our sample tend to have higher $|B_{\ell}|$.  However it should be noted that there are only 5 K dwarfs in the combined set of planet-hosting stars.

\begin{figure}
  \includegraphics[scale=0.45]{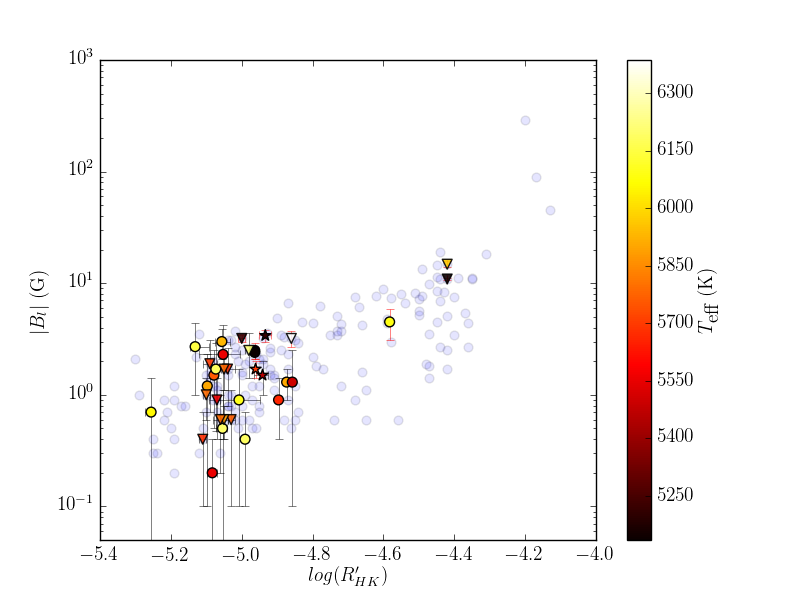}
  \caption{Plot of maximum measured $|B_{\ell}|$ vs $\log(R^{\prime}_{HK})$ (Table~\protect\ref{tab:derivedValues}) for the planet-hosting stars. Symbols are as in Fig.~\protect\ref{fig:HR}.  The complete BCool sample is shown as blue circles.}
  \label{fig:BvsRHK}
\end{figure}

Chromospheric activity seems to be the most strongly correlated with $|B_{\ell}|$, as shown in Figure~\ref{fig:BvsRHK}.  It should be noted that more active stars are generally excluded from planet-search programs, given that phenomena generated by stellar activity can mimic the cyclic variations used to detect planets \citep{Saar1997,Saar1998,Jeffers2014}, additionally biasing the entire sample to less active stars.  Much work is currently underway in attempting to disentangle stellar activity signals from those generated from planetary sources \citep{Petit2015,Feng2016,Hebrard2016,Herrero2016}.  This may in the future allow for a broadening of the sample set of planet-hosting stars. 

\subsection{Planetary Influences}

\begin{figure}
  \includegraphics[scale=0.45]{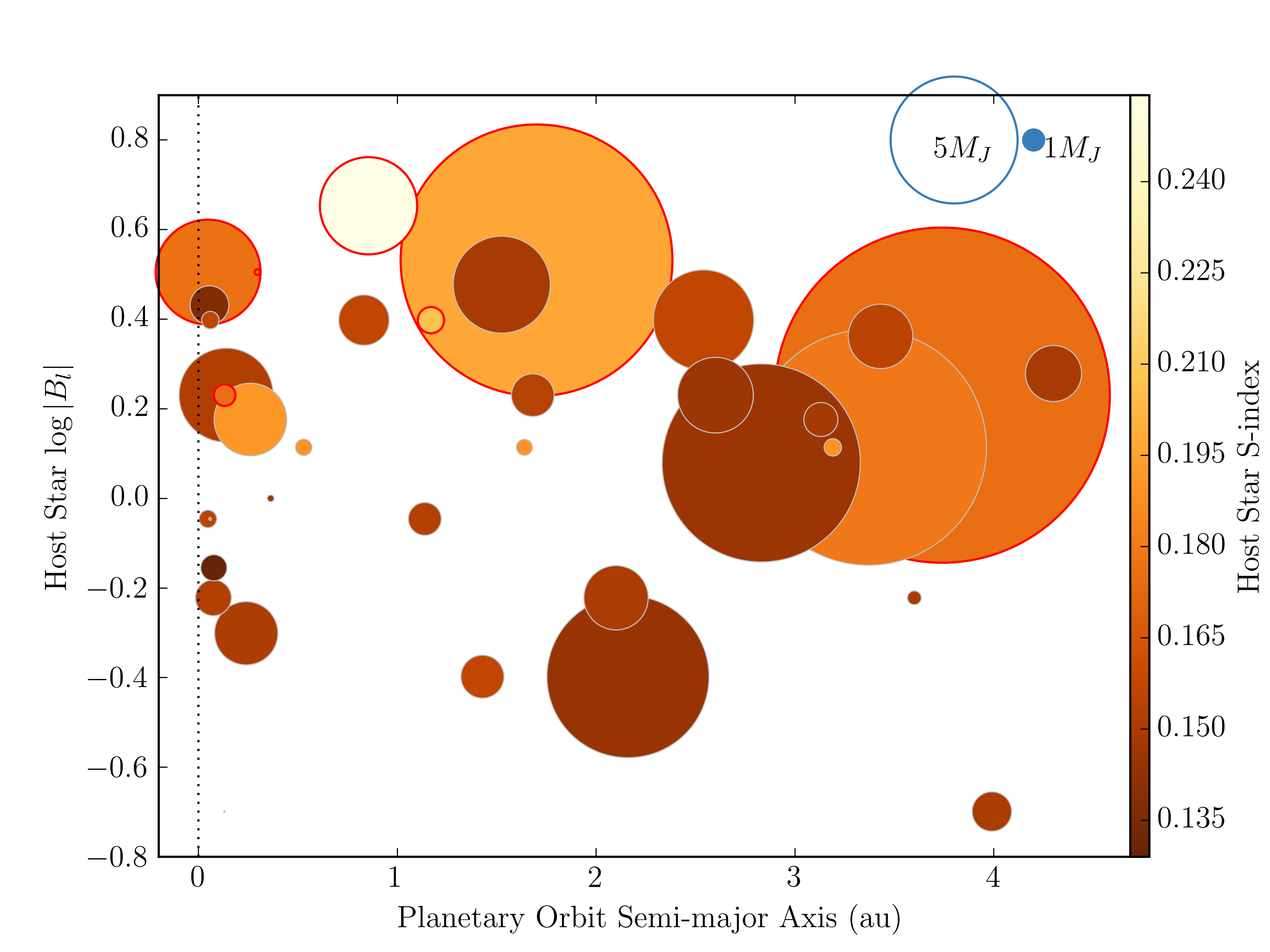}
   \includegraphics[scale=0.45]{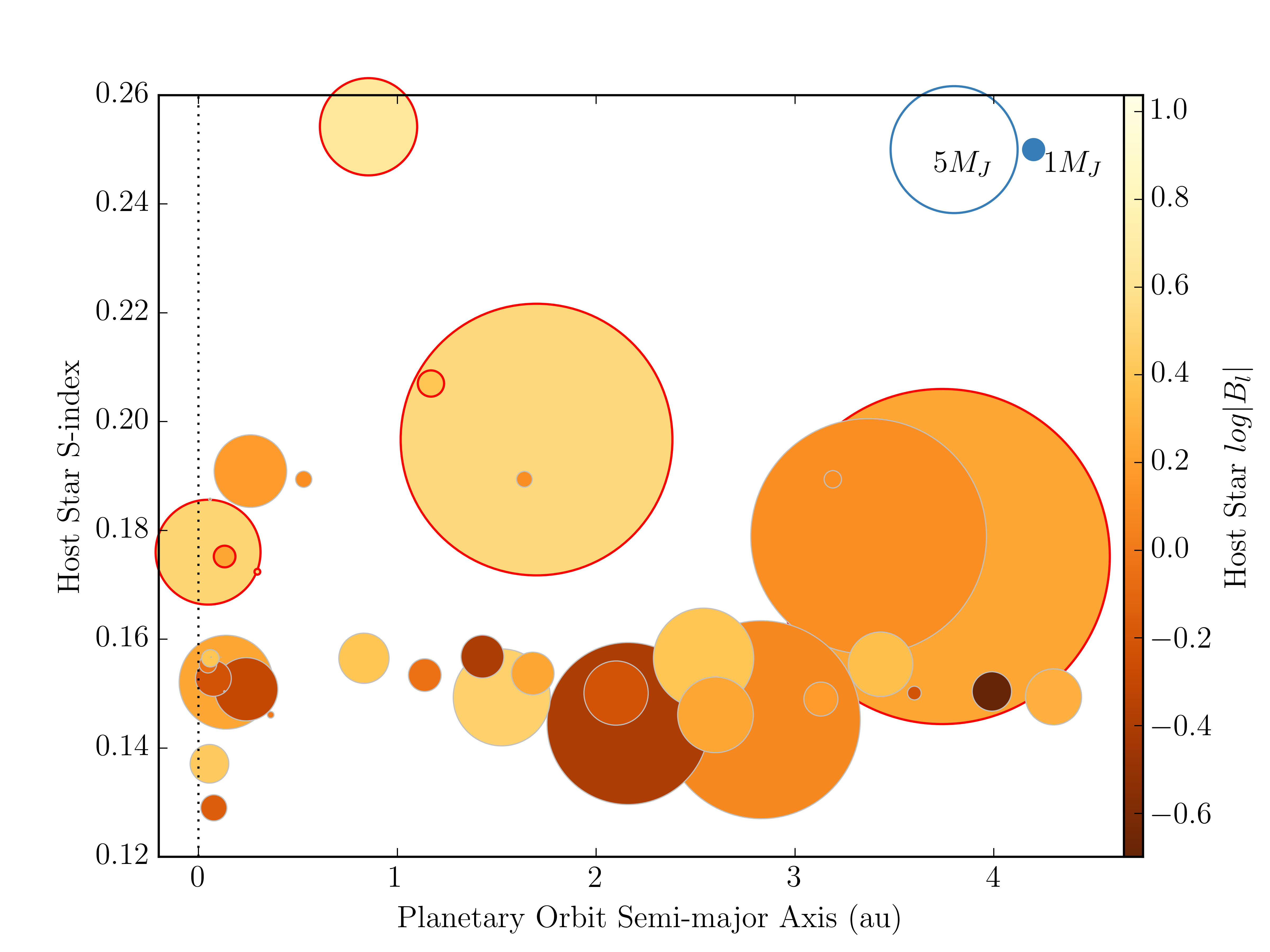}
  \caption{S-index and $\log(|B_{\ell}|)$ shown against the semi-major axis of planetary orbits for the BCool planet hosting stars and the new survey targets.  The fill colours of the circles represent $\log(|B_{\ell}|)$ (upper panel) and S-index (lower panel).  The radius of circles are proportional to planetary masses, $M~\sin~i$. Red edges to the circle indicates the host star had a magnetic detection.  HIP107350 (HN~Peg) and HIP16537 ($\epsilon$~Eri) are not included on this plot for clarity.  HN~Peg~b has a semi-major axis of $\sim \SI{790}{\au}$, and $\epsilon$~Eri has a relatively high S-index (0.5357) compared to the other sample stars; both have magnetic detections and $\log(|B_{\ell}|)$ much higher than all other stars shown.  We find no detectable correlation between mass and orbital distance of the planets in the system and the activity of the star.}
  \label{fig:planet}
\end{figure}

In Figure~\ref{fig:planet}, we present the configuration of the planetary systems from both our new sample and the BCool sample, and their relationship with the activity proxy (S-index) and the magnetic field [$\log(|B_{\ell}|$)].  In both panels, each circle represents a planet, and its centre point on the x-axis represents the semi-major axis of its orbit.  In the case of systems with more than one planet, each planet is represented, with all planets in a system having the same y-axis value. The size of the circles is proportional to the planetary mass.  In the upper panel, the y-axis represents the magnetic field strength of the host star and the colour of the circle its S-index.  In the lower panel, the y-axis represents the S-index of the host star and the colour of the circle is $\log(|B_{\ell}|$).  Finally if a magnetic detection occurred for a host star, the circle(s) representing its planet(s) are outlined in red. There appears to be no correlation between the magnetic field strength and the planetary configuration.  

There are several planets which would warrant the colloquial appellation of ``hot Jupiter'', and would therefore seem the most likely candidates for star-planet interaction, whether through tidal effects, magnetic interaction, or other posited SPI mechanisms.  However, the presence of these planets do not result in any detectable trend towards higher activity or higher measured longitudinal magnetic field.  This is consistent with the findings of \citet{Miller2015}.

\begin{figure}
  \includegraphics[width=\columnwidth]{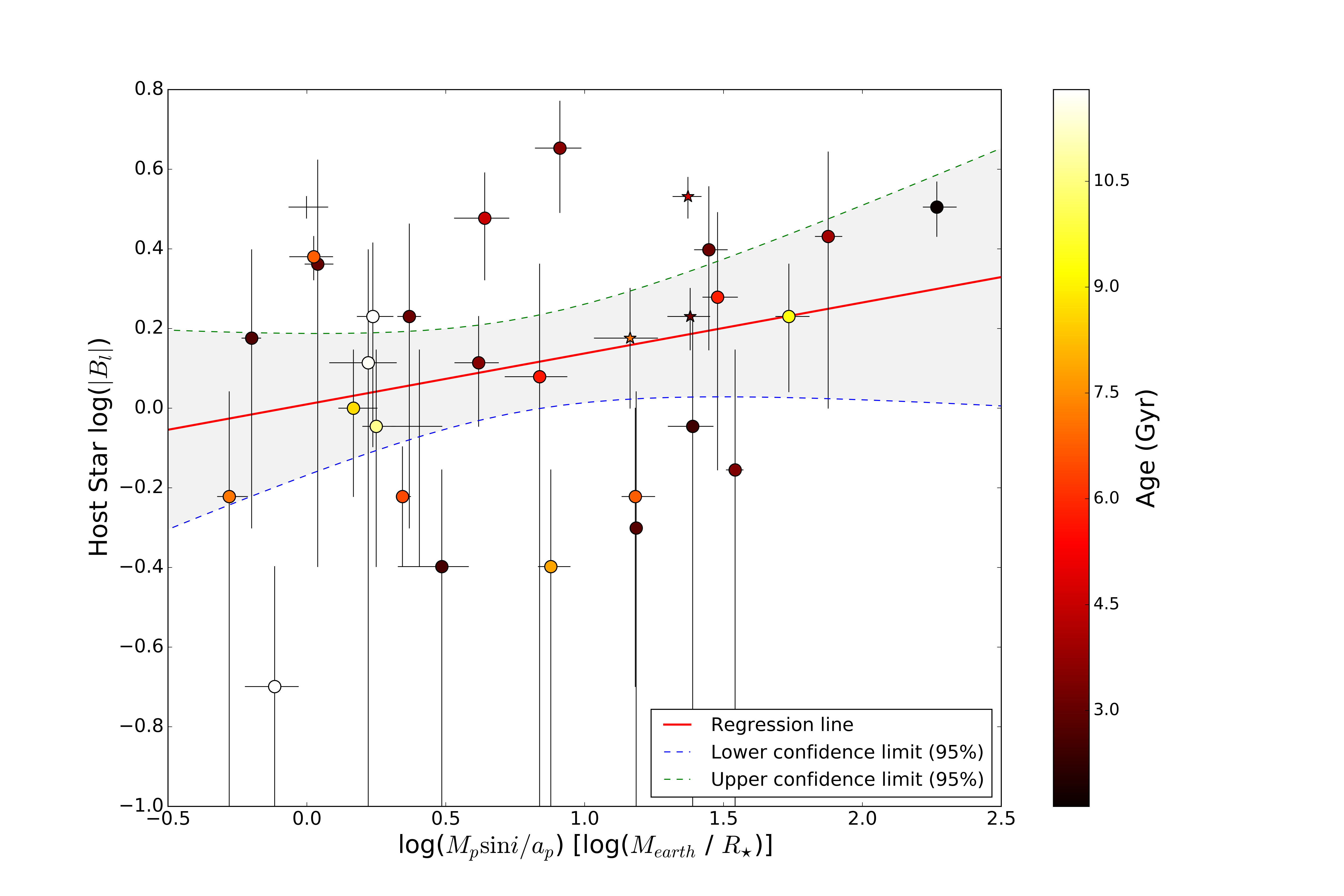}
  \caption{$\log(|B_{\ell}|)$ shown against mass ratio for the BCool planet hosting stars and the new survey targets.  Points are coloured according to age.  Note there is a very weak positive linear relationship (see regression line).  The shaded area represents the 95\% confidence interval.  Within this confidence interval, the population relationship may be zero. The young, active stars HIP107350 (HN Peg) and HIP16537 ($\epsilon$~Eri) are not included on this plot for clarity.}
  \label{fig:massratio}
\end{figure}

\begin{figure}
  \includegraphics[width=\columnwidth]{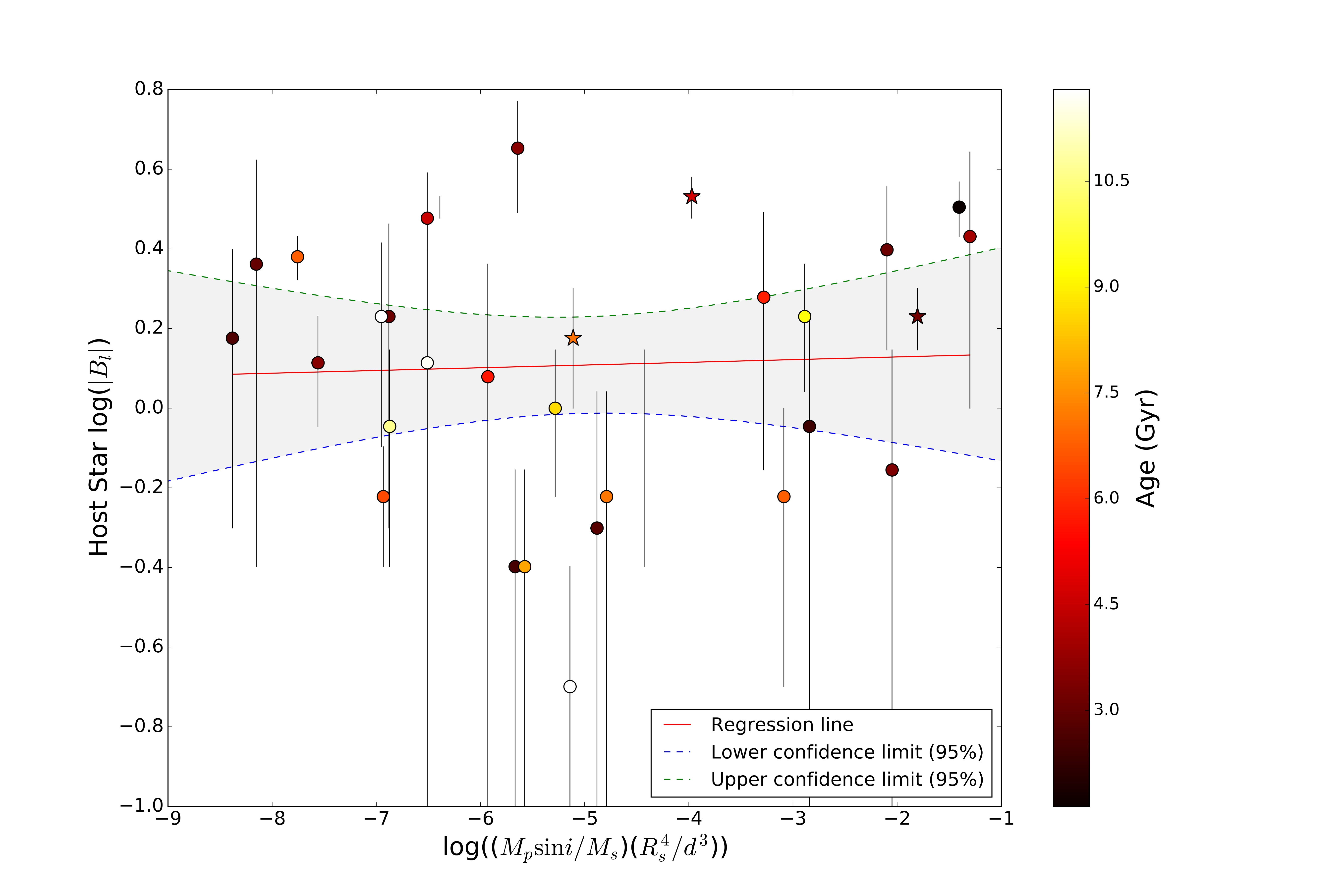}
  \caption{$\log(|B_{\ell}|)$ shown against the log of the relative tidal height ($h_{t}$) for the BCool planet hosting stars and the new survey targets.  Points are coloured according to age.  Horizontal error bars have been omitted for clarity.  No trend is apparent, although a regression line is shown.  The shaded area represents the 95\% confidence interval around the regression line. The young, active stars HIP107350 (HN Peg) and HIP16537 ($\epsilon$~Eri) are not included on this plot for clarity.}
  \label{fig:tidal}
\end{figure}

Indeed investigating potential tidal effects in particular, if we plot $\log(|B_{\ell}|)$ versus the logarithm of the mass ratio (planetary mass divided by the orbital semi-major axis), shown in Figure~\ref{fig:massratio}, there appears to be a weak positive linear relationship.  This should be interpreted with some caution, given the very low $R^{2}$ value and the 95\% confidence interval.  This confidence interval provides an estimate of the uncertainty around the proposed relationship, and indicates that the population relationship may be zero.  Plotting $\log(|B_{\ell}|)$ against the logarithm of the relative height of the tidal bulge induced on the star ($h_{t}$ from \citet[Eq.~1]{Figueira2016}) shows absolutely no correlation (Fig.~\ref{fig:tidal}).


It is noted that the sample is small, and while there are some ``hot Jupiters'' none of them are at the higher end of the mass range, nor are the more massive planets particularly close to the host star.  Given the correlations between stellar parameters and the strength of the stellar magnetic field detailed by \citet{Marsden2014}, it is clear that any effect of SPI on the global stellar magnetic field or activity proxies measured here, if it exists, is too subtle to detect in our sample.  \citet{Fares2013} and \citet{Vidotto2014} came to similar conclusions and called for larger samples.  However, the inherent biases in planet-search programs which exclude active stars makes increasing the sample size and scope difficult.  Alternatively, in the absence of a larger sample, increasing the amount of observations of both the set of planet-hosting and non-planet-hosting stars would allow us to analyse the full magnetic variability range for each star and to look for influence on any magnetic cycles.

\section{Conclusions}

We have presented an investigation of the large-scale magnetic field of 19 additional solar-type stars for the BCool survey \citep{Marsden2014}.  We expanded our sample of planet-hosting stars by adding the previously-observed planet host in the BCool survey.  The results we obtain for these stars are congruent with the wider BCool survey.  The selection by planet-search surveys of low activity, mature stars biases the sample of stars with discovered planets towards those with higher ages and lower intrinsic activity.  Consequently, we obtain a lower rate of magnetic detections, and of lower measures of $|B_{\ell}|$, the longitudinal magnetic field, than the wider survey of 170 solar-type stars.

While we cannot rule out that the presence of the planets around these host stars has an effect on the host star's magnetic field, we show that such an effect is too subtle to detect in our sample.  A larger sample of planet-hosting solar-type stars may reveal a trend.  If a trend exists, there may be a lower limit to the planetary mass and upper limit to the semi-major axis of the planetary orbit for which any SPI becomes apparent using these methods.  Further target selection should perhaps focus on massive hot Jupiter systems to further investigate these limits.

The four planetary systems for which we have made magnetic detections may be candidates for long-term magnetic topology monitoring (see Appendix~\ref{sec:appA}).  Observation of the magnetic field of the host star may allow for modelling of the space weather environment of its planetary system \citep{Vidotto2015,Nicholson2016,Alvarado2016}.  Also, with these known magnetic fields, it may be possible to determine whether other methods, such as radio observations, may be able to detect magnetic interactions between star and planet, or to at least place limits on the expected signals one may expect from such behaviour \citep{Fares2010,Vidotto2012,See2015,Vidotto2015}.

\section*{Acknowledgments}

This work was based on observations obtained with NARVAL at the T\'elescope Bernard Lyot (TBL). TBL/NARVAL are operated by INSU/CNRS.

SVJ acknowledges research funding by the Deutsche Forschungsgemeinschaft (DFG) under grant SFB 963/1, project A16.


The Strategic Research Funding for the Starwinds project provided by the University of Southern Queensland provides resources to the Astrophysics group within the Computational Engineering and Science Research Centre at USQ (MWM, SCM, BDC).  MWM is supported by an Australian Postgraduate Award Scholarship.

This research has made use of NASA's Astrophysics Data System. This research has made use of the SIMBAD database, operated at CDS, Strasbourg, France. This research has made use of the VizieR catalogue access tool, CDS, Strasbourg, France.  This work has made use of the VALD database, operated at Uppsala University, the Institute of Astronomy RAS in Moscow, and the University of Vienna.

The authors would like to thank Victor See and Colin Folsom for their helpful comments and suggestions on the original manuscript.  Further thanks to the BCool collaboration.


\bibliography{paper.bib}
\bibliographystyle{mnras}

\appendix

\section[]{Magnetic Detections}
\label{sec:appA}

\subsection{HIP~27253 (HD~38529)}

\begin{figure}
  \includegraphics[scale=0.45]{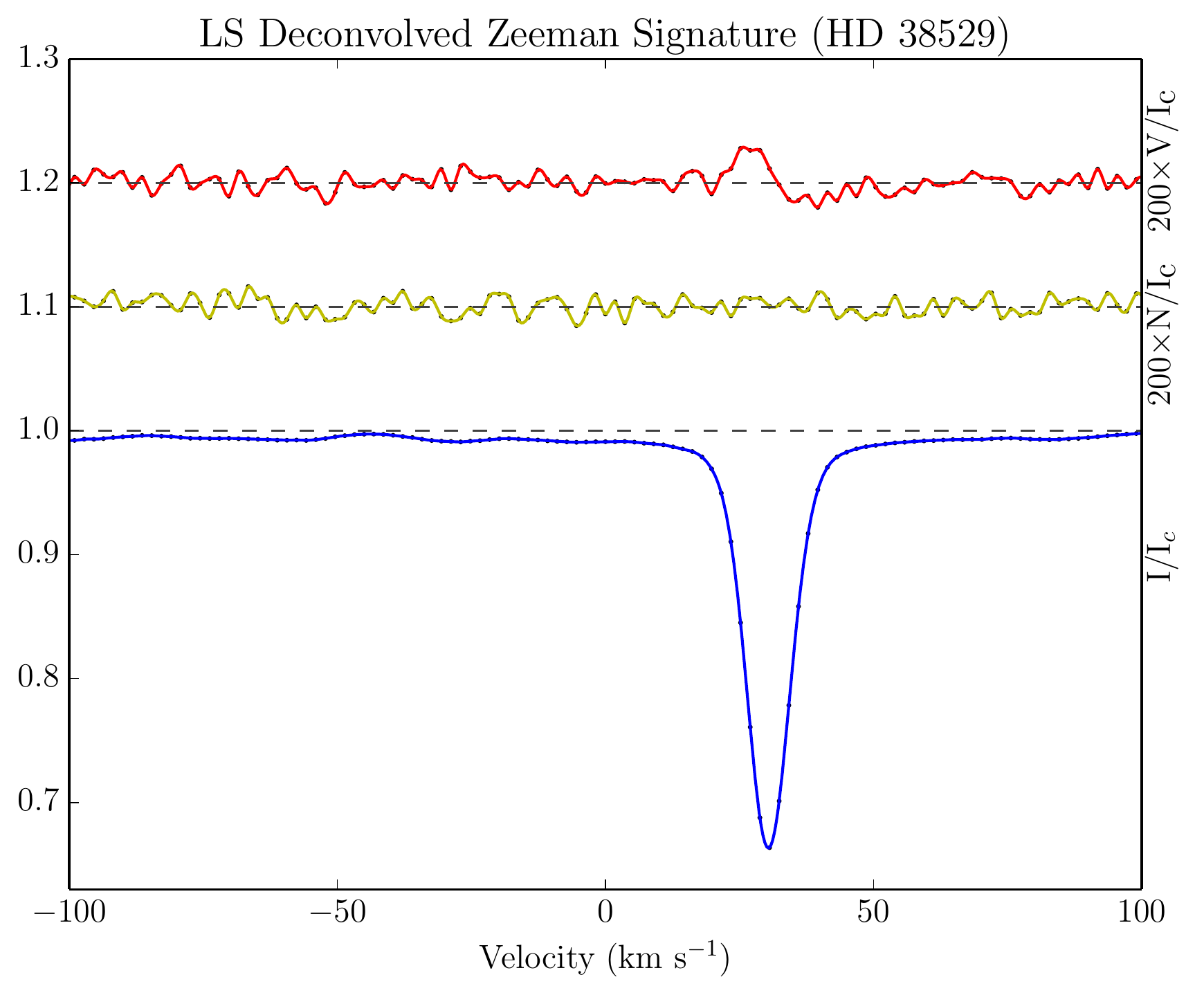}
  \caption{Plot of the LSD profiles of HIP 27253 (HD 38529).  The upper line (in red) is the Stokes $V$ profile, expanded 200 times and shifted up by 0.2.  The centre line (yellow) represents the LSD null profile also expanded by 200 and shifted up 0.1.  The lower line (in blue) shows the Stokes $I$ (intensity) LSD profile.}
  \label{fig:HIP27253}
\end{figure}

The star HIP~27253 (G8III/IV; $T_{eff} = \SI[separate-uncertainty=true,multi-part-units=single]{5697 \pm 44}{\kelvin}$) is known to host two planets.  HD~38529b has a minimum mass $\sim0.85~M_J$, and orbits at a distance of $\sim\SI{0.13}{\au}$, while the much more massive HD~38529c ($\sim13~M_J$) orbits at approximately \SI{3.75}{\au}.  HIP~27253 has the lowest S-index of the four stars for which we obtained detections ($0.1750\pm0.0005$).  It is relatively bright ($V\sim6$) compared with the remainder of the sample, with over 12000 lines used in the LSD process, resulting in a relatively high $\mbox{SNR}_{\textsc{LSD}}$ of $\sim28000$.

One of the two subgiants for which we obtained a magnetic detection, \citet{Takeda2007} place its age at over \SI{3}{\giga\year}, whilst we derive a chromospheric age of $\sim\SI{4.8}{\giga\year}$.  An older subgiant star with a close Jupiter mass planet, HIP~27253 would be an interesting target for further investigation.  The LSD profile including Stokes $V$, Stokes $I$ and null ($N$) profiles is shown in Figure~\ref{fig:HIP27253}.  

\subsection{HIP~27384 (HD~38801)}
\label{sec:HIP27384}

\begin{figure}
  \includegraphics[scale=0.45]{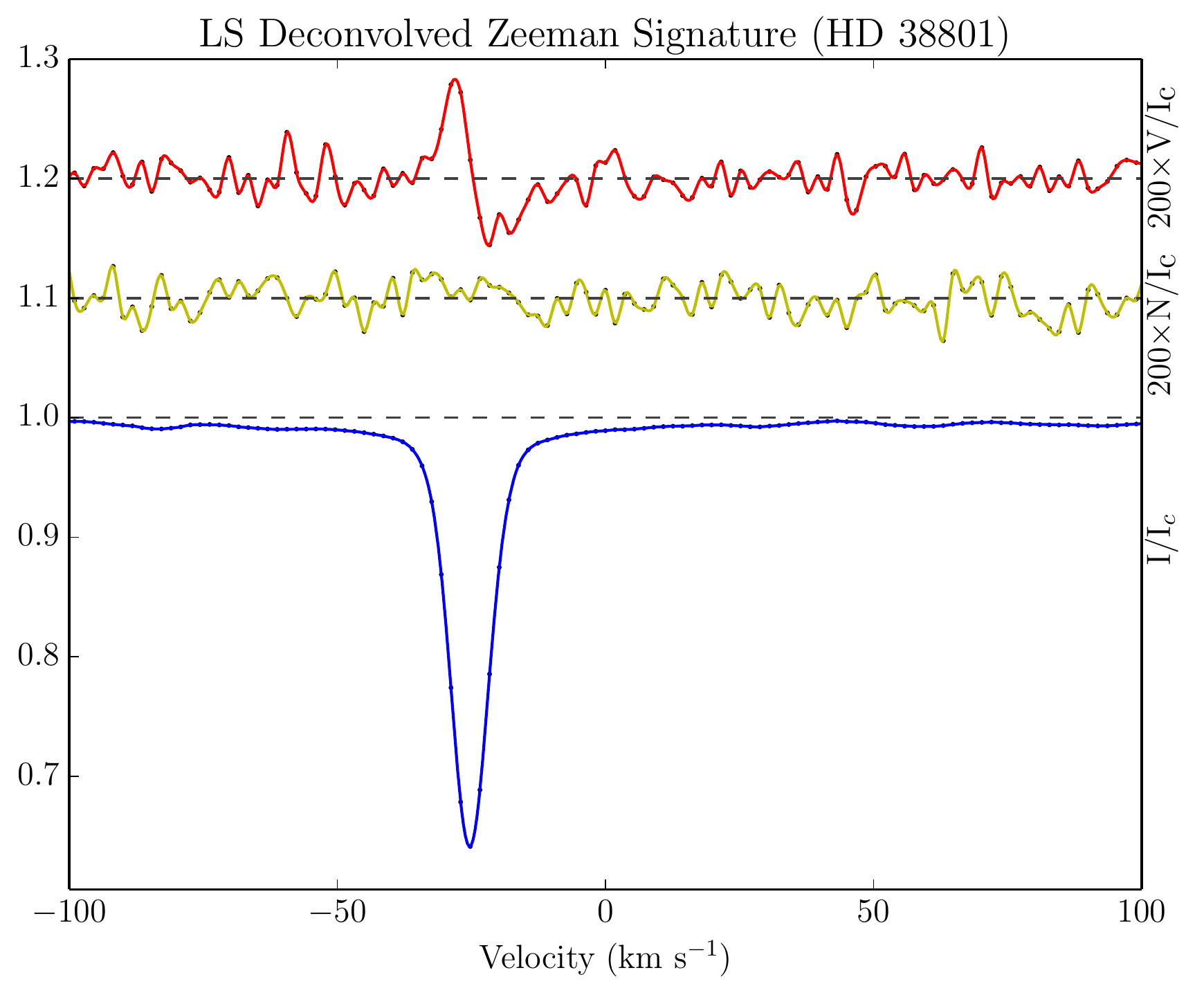}
  \caption{Plot of LSD profiles for HIP~27384 (HD~38801).  The plot is described in Figure~\protect\ref{fig:HIP27253}.}
  \label{fig:HIP27384}
\end{figure}

HIP~27384 (G8IV; $T_{eff} = \SI[separate-uncertainty=true,multi-part-units=single]{5222 \pm 44}{\kelvin}$) is the second of our two subgiants which have a magnetic detection (LSD profiles shown in Fig.~\ref{fig:HIP27384}).  A single large planet HD~38801b with a mass of $\sim10.7~M_J$ orbits the star at \SI{1.7}{\au}.  A mature star with age measurements $>\SI{4}{\giga\year}$, HIP~27384 has one of the largest $|B_{\ell}|$ measurements in our sample.  Its low $v \sin i \approx \SI{0.5}{\kilo\meter\per\second}$ makes tomographic mapping of the surface magnetic field challenging.

\subsection{HIP~28767 (HD~40979)}
\label{sec:HIP28767}

\begin{figure}
  \includegraphics[scale=0.45]{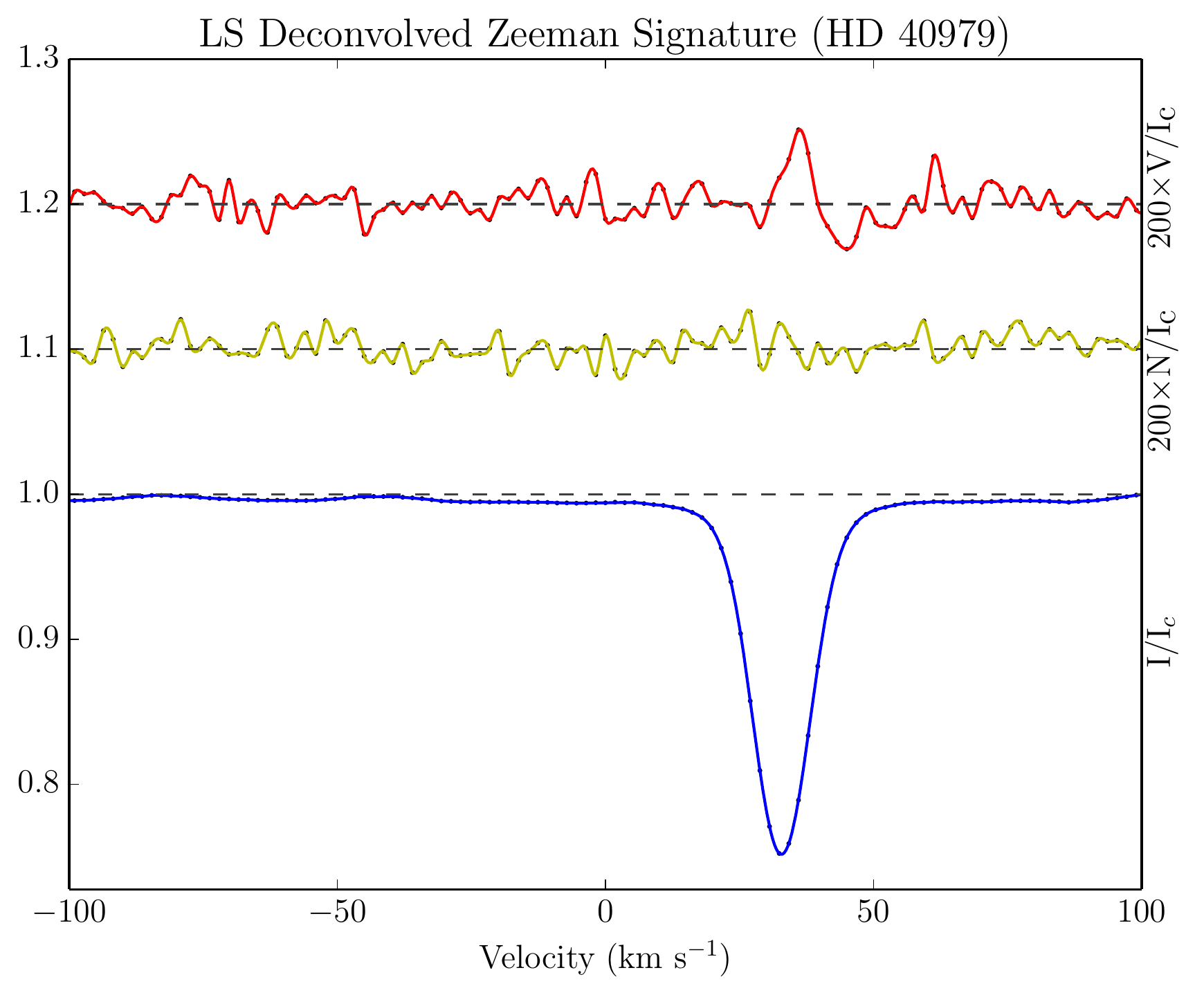}
  \caption{Plot of LSD profiles for HIP 28767 (HD 40979).  The plot is described in Figure~\protect\ref{fig:HIP27253}.}
  \label{fig:HIP28767}
\end{figure}

HIP~28767 (F8; $T_{eff} = \SI[separate-uncertainty=true,multi-part-units=single]{6089 \pm 44}{\kelvin}$) has by far the highest S-index of the sample ($0.2543 \pm 0.0003$), and consequently it is not surprising that a magnetic field was detected.  A significant field in the null profile ($|N_{\ell}|$) relative to the size of the measured $|B_{\ell}|$ means that followup observations may be necessary to confirm the actual level of the longitudinal field with more confidence.  The hottest star in our sample to have a detection, HIP~28767 has a $\sim3.8~M_J$ planet (HD~40979b) orbiting at a distance of $~\sim\SI{0.85}{\au}$.  The Stokes $V$, Stokes $I$ and $N$ profiles for HIP~28767 are shown in Figure~\ref{fig:HIP28767}.

\subsection{HIP~64457 (HD~114783)}

\begin{figure}
  \includegraphics[scale=0.45]{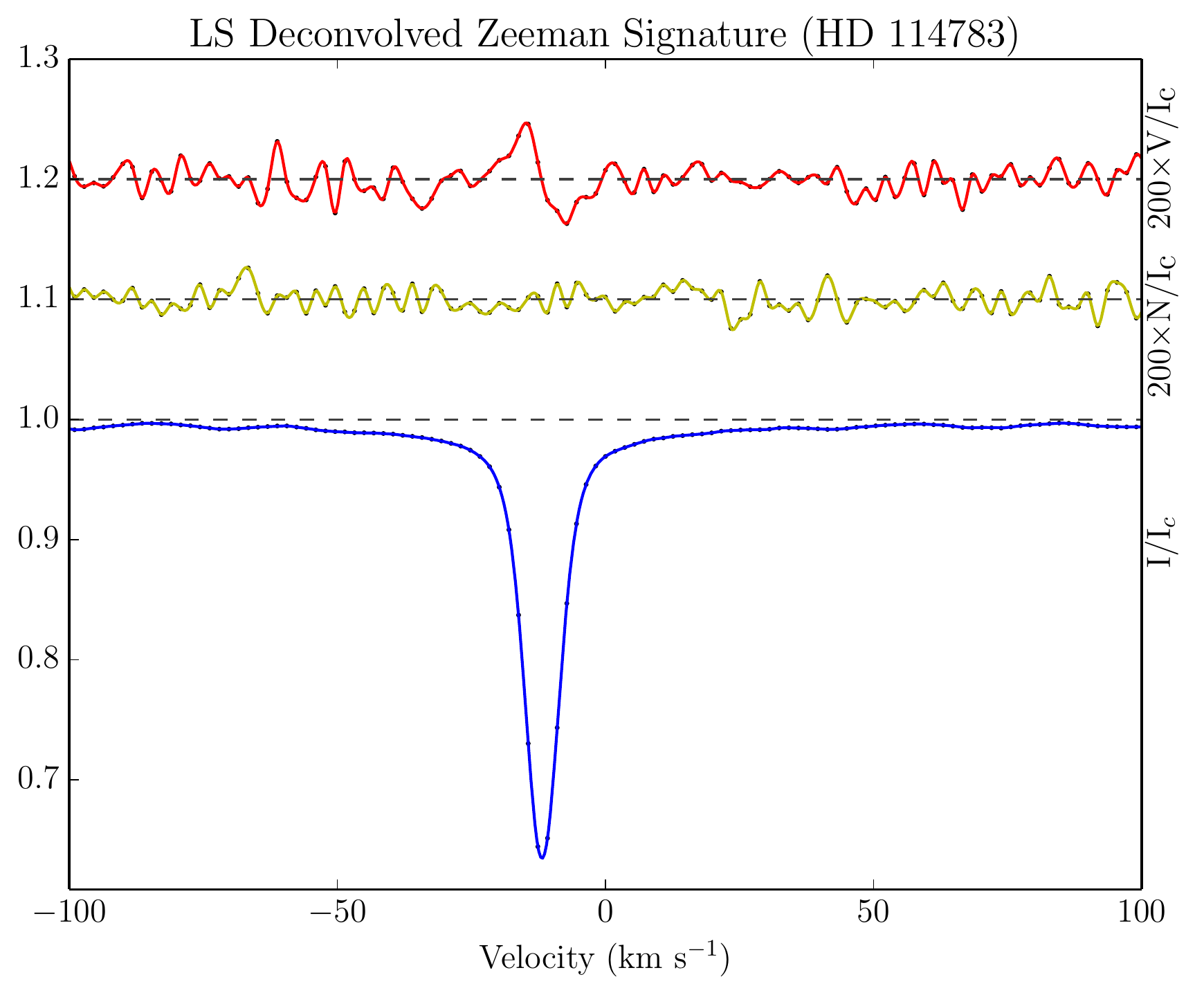}
  \caption{Plot of LSD profiles for HIP~64457 (HD~114783).  The plot is described in Figure~\protect\ref{fig:HIP27253}.}
  \label{fig:HIP64457}
\end{figure}

Two observations of HIP~64457 (K1V; $T_{eff} = \SI[separate-uncertainty=true,multi-part-units=single]{5135 \pm 44}{\kelvin}$) were taken during the observing period, and the set of LSD profiles from the second observation (taken at $\approx$~HJD~2457032.7) is shown in Figure~\ref{fig:HIP64457}.  HD~114783b, with a mass approximately  that of Jupiter orbits the star at a distance of $\sim\SI{1.1}{\au}$.  As one of the two stars (the other being HIP~28767; Sec.~\ref{sec:HIP28767})  with an S-index above 0.2, it was likely to provide a magnetic detection.  With a similar temperature and rotation rate, a comparison of the magnetic activity and/or cycles of HIP~64457 and HIP~27384 (Sec.~\ref{sec:HIP27384}) may provide an insight into any apparent differences or similarities of dynamo between dwarfs and subgiants.

\section[]{Planet Hosts from BCool Survey}
\label{BCappendix}

\begin{table*}
\caption{Planetary parameters of the sample of planet-hosting solar type stars from the BCool survey \protect\citep{Marsden2014}, as an extension of Table~\protect\ref{tab:planetParams}.  The stellar component's \textit{Hipparcos} number, SPOCS catalogue number and HD number (where applicable) are shown in the first three columns.  Column 4 refers to the Name by which the planetary components (column 5) are known.  The period, mass ($M \sin i$) and semi-major axis is shown for each planet.  These values are from the references listed in the last column.}
\label{tab:planetParamsBC}
\begin{tabular}{lccccccccc}
\hline
\multicolumn{3}{c}{Star} &  & \multicolumn{6}{c}{Planet(s)} \\\cline{1-3}\cline{5-10}

\multicolumn{1}{c}{HIP} & \multicolumn{1}{c}{SPOCS} & \multicolumn{1}{c}{HD}  &  & \multicolumn{1}{c}{Name} & \multicolumn{1}{c}{Component} &  \multicolumn{1}{c}{Period} & \multicolumn{1}{c}{$M \sin i$} & \multicolumn{1}{c}{Semi-major} & \multicolumn{1}{c}{Refs.}  \\
\multicolumn{1}{c}{no.} & \multicolumn{1}{c}{no.}  & \multicolumn{1}{c}{no.} & &  \multicolumn{1}{c}{} & \multicolumn{1}{c}{} & \multicolumn{1}{c}{(\SI{}{\day})} & \multicolumn{1}{c}{($M_{J}$)} & \multicolumn{1}{c}{axis (\SI{}{\au})} & \multicolumn{1}{c}{}  \\
\hline \\[-1.5ex]
1499 & 13 & 1461 & & HD~1461 & b & $5.77152 \pm 0.00045$ & $0.0203 \pm 0.0019$ & $0.0634 \pm 0.0022$ & 2\\[3pt]
~ & ~ & ~ & & HD~1461 & c & $13.5052 \pm 0.0029$ & $0.0176 \pm 0.0023$ & $0.1117 \pm 0.0039$ & 2\\[3pt]
3093 & 26 & 3651 & & 54~Psc & b & $62.206 \pm 0.021$ & $0.227 \pm 0.023$ & $0.296 \pm 0.017$ & 1\\[3pt]

7513 & 85 & 9826 & & $\upsilon$~And & b & $4.617113 \pm 0.000082$ & $0.687 \pm 0.058$ & $0.0595 \pm 0.0034$ & 1\\[3pt]
~ & ~ & ~ & & $\upsilon$~And & c & $241.23 \pm 0.30$ & $1.98 \pm 0.17$ & $0.832 \pm 0.048$ & 1\\[3pt]
~ & ~ & ~ & & $\upsilon$~And & d & $1290.1 \pm 8.4$ & $3.95 \pm 0.33$ & $2.54 \pm 0.15$ & 1\\[3pt]

8159 & 97 & 10697 & & HD~10697 & b & $1076.4 \pm 2.4$ & $6.38 \pm 0.53$ & $2.16 \pm 0.12$ & 1\\[3pt]

12048 & 128 & 16141 & & HD~16141 & b & $75.523 \pm 0.055$ & $0.260 \pm 0.028$ & $0.363 \pm 0.021$ & 1\\[3pt]

16537$^{+}$ & 171 & 22049 & & $\epsilon$~Eri & b & $2502 \pm 10$ & $0.78 \pm 0.08$ & $3.39 \pm 0.36$ & 3\\[3pt]

53721 & 472 & 95128 & & 47~Uma & b & $1078^{+2}_{-2}$ & $2.53^{+0.07}_{-0.06}$ & $2.100^{+0.02}_{-0.02}$ & 4\\[3pt]
~ & ~ & ~ & & 47~Uma & c & $2391^{+100}_{-87}$ & $0.540^{+0.066}_{-0.073}$ & $3.6^{+0.1}_{-0.1}$ & 4\\[3pt]
~ & ~ & ~ & & 47~Uma & d & $14002^{+4018}_{-5095}$ & $1.64^{+0.29}_{-0.48}$ & $11.6^{+2.1}_{-2.9}$ & 4\\[3pt]

67275 & 577 & 120136 & & $\tau$~Bo\"o & b & $3.312463 \pm 0.000014$ & $4.13 \pm 0.34$ & $0.0481 \pm 0.0028$ & 1\\[3pt]

96901 & 855 & 186427 & & 16~Cyg~B & b & $798.5 \pm 1.0$ & $1.68 \pm 0.15$ & $1.681 \pm 0.097$ & 1\\[3pt]

100970 & 894 & 195019 & & HD~195019 & b & $18.20163 \pm 0.00040$ & $3.70 \pm 0.30$ & $0.1388 \pm 0.0080$ & 5\\[3pt]

107350$^{*}$ & 942 & 206860 & & HN~Peg & b & ~ & $16.0 \pm 9.4$ & $795.0 \pm 15.0$ & 6\\[3pt]

109378 & 960 & 210277 & & HD~210277 & b & $442.19 \pm 0.50$ & $1.29 \pm 0.11$ & $1.138 \pm 0.066$ & 1\\[3pt]

113357 & 990 & 217014 & & 51~Peg & b & $4.230785 \pm 0.000036$ & $0.472 \pm 0.039$ & $0.0527 \pm 0.0030$ & 1\\[3pt]

113421 & 992 & 217107 & & HD~217107 & b & $7.12690 \pm 0.00022$ & $1.41 \pm 0.12$ & $0.0748 \pm 0.0043$ & 1\\[3pt]
~ & ~ & ~ & & HD~217107 & c & $3200 \pm 1000$ & $2.21 \pm 0.66$ & $4.3 \pm 1.2$ & 1\\[3pt]

\hline
\end{tabular} \\
References: 1: \cite{Butler2006}, 2: \cite{Diaz2016}, 3: \cite{Benedict2006}, 4: \cite{Gregory2010}, 5: \cite{Wright2007}, 6: \cite{Luhman2007}. \\

$^{+}$  -  HIP16537 ($\epsilon$~Eri) is suspected to host a second planet \citep{Quillen2002}. \\
$^{*}$  -  The planet orbiting HIP107530 (HN~Peg) was discovered by direct imaging, and as a result, its orbital parameters unclear; semi-major axis is therefore taken as the projected separation \citep{Luhman2007}.
\end{table*}

\begin{table*}
\scriptsize
\begin{center}
\caption{Stellar parameters of the sample of planet-hosting solar type stars from the BCool survey \protect\citep[Table~1]{Marsden2014}, as an extension of Table~\protect\ref{tab:stellarParams}.  The spectral type is from SIMBAD (\url{http://simbad.u-strasbg.fr/simbad/}, \protect\citet{Wenger2000}).  Column 11 is the radius of the convective zone of the star. Values are found in the references shown in the final column of the table; locations where values were unavailable in the literature, are denoted by `$X$'.}
\label{tab:stellarParamsBC}
\begin{tabular}{lccccrccccccc}
\hline
\multicolumn{1}{c}{HIP} & \multicolumn{1}{c}{SPOCS} & \multicolumn{1}{c}{Spec.} & \multicolumn{1}{c}{$T_{\mbox{eff}}$} & \multicolumn{1}{c}{log($g$)} & \multicolumn{1}{c}{[M/H]} & \multicolumn{1}{c}{log(Lum)} & \multicolumn{1}{c}{Age} & \multicolumn{1}{c}{Mass} & \multicolumn{1}{c}{Radius} & \multicolumn{1}{c}{Radius$_{\mbox{CZ}}$} & \multicolumn{1}{c}{$v \sin i$} & \multicolumn{1}{c}{Refs.} \\
\multicolumn{1}{c}{no.} & \multicolumn{1}{c}{no.}  & \multicolumn{1}{c}{Type} & \multicolumn{1}{c}{(\SI{}{\kelvin})} & \multicolumn{1}{c}{(\SI{}{\centi\meter\per\second\squared})} & \multicolumn{1}{c}{} & \multicolumn{1}{c}{($L_{\sun}$)} & \multicolumn{1}{c}{(\SI{}{\giga\year})} & \multicolumn{1}{c}{($M_{\sun}$)} & \multicolumn{1}{c}{($R_{\sun}$)} & \multicolumn{1}{c}{($R_{\sun}$)} & \multicolumn{1}{c}{(\SI{}{\kilo\meter\per\second})} & \\
\hline \\[-1.5ex]
1499 & 13 & G0V & 5765$^{+44}_{-44}$ & 4.37$^{+0.03}_{-0.03}$ & +0.16$^{+0.03}_{-0.03}$ & +0.078$^{+0.041}_{-0.041}$ & 7.12$^{+1.40}_{-1.56}$ & 1.026$^{+0.040}_{-0.030}$ & 1.11$^{+0.04}_{-0.04}$ & 0.323$^{+0.019}_{-0.020}$ & 1.6$^{+0.5}_{-0.5}$ & 1,2 \\[3pt]

3093 & 26 & K0V & 5221$^{+25}_{-25}$ & 4.51$^{+0.02}_{-0.01}$ & +0.16$^{+0.02}_{-0.02}$ & -0.286$^{+0.018}_{-0.018}$ & $X$ & 0.882$^{+0.026}_{-0.021}$ & 0.88$^{+0.03}_{-0.02}$ & 0.296$^{+0.006}_{-0.009}$ & 1.1$^{+0.3}_{-0.3}$ & 1,2 \\[3pt]

7513 & 85 & F9V & 6213$^{+22}_{-22}$ & 4.16$^{+0.02}_{-0.04}$ & +0.12$^{+0.01}_{-0.01}$ & +0.522$^{+0.021}_{-0.021}$ & 3.12$^{+0.20}_{-0.24}$ & 1.310$^{+0.021}_{-0.014}$ & 1.64$^{+0.04}_{-0.05}$ & 0.315$^{+0.028}_{-0.073}$ & 9.6$^{+0.3}_{-0.3}$ & 1,2 \\[3pt]

8159 & 97 & G5IV & 5680$^{+44}_{-44}$ & 4.03$^{+0.03}_{-0.03}$ & +0.10$^{+0.03}_{-0.03}$ & +0.448$^{+0.054}_{-0.054}$ & 7.84$^{+0.40}_{-0.48}$ & 1.112$^{+0.026}_{-0.020}$ & 1.73$^{+0.06}_{-0.07}$ & 0.568$^{+0.038}_{-0.032}$ & 2.5$^{+0.5}_{-0.5}$ & 1,2 \\[3pt]

12048 & 128 & G5IV & 5794$^{+44}_{-44}$ & 4.19$^{+0.04}_{-0.04}$ & +0.09$^{+0.03}_{-0.03}$ & +0.30$^{+0.10}_{-0.10}$ & 8.68$^{+0.76}_{-0.76}$ & 1.052$^{+0.026}_{-0.022}$ & 1.39$^{+0.07}_{-0.07}$ & 0.416$^{+0.030}_{-0.028}$ & 1.9$^{+0.5}_{-0.5}$ & 1,2 \\[3pt]

16537 & 171 & K2Vk: & 5146$^{+31}_{-31}$ & 4.61$^{+X}_{-0.02}$ & +0.00$^{+0.02}_{-0.02}$ & -0.486$^{+0.011}_{-0.011}$ & 0.00$^{+0.60}_{-0.00}$ & 0.856$^{+0.006}_{-0.008}$ & 0.77$^{+0.02}_{-0.01}$ & 0.235$^{+0.005}_{-0.006}$ & 2.4$^{+0.4}_{-0.4}$ & 1,2 \\[1mm]

53721 & 472 & G1V & 5882$^{+16}_{-16}$ & 4.31$^{+0.03}_{-0.04}$ & +0.02$^{+0.01}_{-0.01}$ & +0.206$^{+0.021}_{-0.021}$ & 6.48$^{+1.44}_{-1.04}$ & 1.063$^{+0.022}_{-0.029}$ & 1.24$^{+0.04}_{-0.04}$ & 0.325$^{+0.028}_{-0.026}$ & 2.8$^{+0.2}_{-0.2}$ & 1,2 \\[3pt]

67275 & 577 & F6IV & 6387$^{+25}_{-25}$ & 4.27$^{+0.04}_{-0.03}$ & +0.25$^{+0.02}_{-0.02}$ & +0.481$^{+0.024}_{-0.024}$ & 1.64$^{+0.44}_{-0.52}$ & 1.341$^{+0.054}_{-0.039}$ & 1.46$^{+0.05}_{-0.05}$ & 0.230$^{+0.010}_{-0.005}$ & 15.0$^{+0.3}_{-0.3}$ & 1,2 \\[3pt]

96901 & 855 & G3V & 5674$^{+17}_{-17}$ & 4.30$^{+0.04}_{-0.02}$ & +0.02$^{+0.01}_{-0.01}$ & +0.095$^{+0.024}_{-0.024}$ & 11.80$^{+2.20}_{-2.00}$ & 0.956$^{+0.026}_{-0.025}$ & 1.17$^{+0.04}_{-0.03}$ & 0.371$^{+0.023}_{-0.024}$ & 2.2$^{+0.2}_{-0.2}$ & 1,2 \\[3pt]

100970 & 894 & G3IV-V & 5788$^{+44}_{-44}$ & 4.18$^{+0.03}_{-0.04}$ & +0.00$^{+0.03}_{-0.03}$ & +0.286$^{+0.069}_{-0.069}$ & 9.32$^{+0.76}_{-0.72}$ & 1.025$^{+0.020}_{-0.018}$ & 1.38$^{+0.06}_{-0.05}$ & 0.413$^{+0.026}_{-0.028}$ & 2.5$^{+0.5}_{-0.5}$ & 1,2 \\[3pt]

107350 & 942 & G0V & 5974$^{+25}_{-25}$ & 4.48$^{+0.01}_{-0.03}$ & -0.01$^{+0.02}_{-0.02}$ & +0.062$^{+0.033}_{-0.033}$ & 0.00$^{+0.88}_{-0.00}$ & 1.103$^{+0.012}_{-0.016}$ & 1.04$^{+0.02}_{-0.03}$ & 0.239$^{+0.010}_{-0.007}$ & 10.6$^{+0.3}_{-0.3}$ & 1,2 \\[3pt]
109378 & 960 & G0 & 5555$^{+44}_{-44}$ & 4.39$^{+0.03}_{-0.03}$ & +0.20$^{+0.03}_{-0.03}$ & -0.020$^{+0.035}_{-0.035}$ & 10.64$^{+2.04}_{-2.20}$ & 0.986$^{+0.038}_{-0.052}$ & 1.06$^{+0.03}_{-0.04}$ & 0.343$^{+0.019}_{-0.022}$ & 1.8$^{+0.5}_{-0.5}$ & 1,2 \\[3pt]

113357 & 990 & G2.5IVa & 5787$^{+25}_{-25}$ & 4.36$^{+0.04}_{-0.03}$ & +0.15$^{+0.02}_{-0.02}$ & +0.117$^{+0.025}_{-0.025}$ & 6.76$^{+1.64}_{-1.48}$ & 1.054$^{+0.039}_{-0.036}$ & 1.15$^{+0.04}_{-0.04}$ & 0.327$^{+0.058}_{-0.024}$ & 2.6$^{+0.3}_{-0.3}$ & 1,2 \\[3pt]
113421 & 992 & G8IV & 5704$^{+44}_{-44}$ & 4.42$^{+0.04}_{-0.03}$ & +0.27$^{+0.03}_{-0.03}$ & +0.050$^{+0.031}_{-0.031}$ & 5.84$^{+1.92}_{-2.44}$ & 1.108$^{+0.034}_{-0.052}$ & 1.08$^{+0.04}_{-0.03}$ & 0.316$^{+0.022}_{-0.018}$ & 0.0$^{+0.5}_{-0.0}$ & 1,2 \\[3pt]

\hline
\end{tabular}
\end{center}
References: 1: \cite{Valenti2005}, 2: \cite{Takeda2007}. \\
\end{table*}

\begin{table*}
\scriptsize
\caption{Results from the analysis of the Stokes $V$ LSD profiles of the planet hosting stars in the BCool sample \protect\citep[Table~3]{Marsden2014}.  Column 3 provides the date of the observation corresponding to the observation number shown in column 2.  Column 4 shows the radial velocity for the star determined by \protect\citep{Marsden2014}; for comparison column 5 shows the radial velocity measured by \protect\citet{Nidever2002} ($^{\textsc{NS}}$indicates a non-radial velocity standard star; $\sigma_{res} \ge \SI{100}{\meter\per\second}$).  Columns 6 and 7 show the signal-to-noise of the Stokes $V$ profile and the number of lines used in the LSD process respectively.  Column 8 indicates is the magnetic field was unambiguously detected (D), marginal (M) or not (N).  Stars indicated by $^{+}$ have multiple observations and where appropriate the number of definite (D) and marginal (M) detections are shown as fractions of the total number of observations.  HIP113357 ($^{*}$ has multiple observations; all non-detections (N).  Column 9 shows the false alarm probability calculated for the detection in column 8.  Columns 10 and 11 indicate the velocity range used to calculate $B_{\ell}$ (column 12) and $N_{\ell}$ (column 13) using equation \protect\ref{eq:Bl}.}
\label{tab:magneticFieldBC}
\begin{tabular}{lcccccccccccccc}
\hline
\multicolumn{1}{c}{HIP} & \multicolumn{1}{c}{Obs.} & \multicolumn{1}{c}{Obs.} & \multicolumn{1}{c}{RV} & \multicolumn{1}{c}{RV} & \multicolumn{1}{c}{SNR$_{\mbox{LSD}}$} & \multicolumn{1}{c}{lines} & \multicolumn{1}{c}{Detection} & \multicolumn{1}{c}{FAP} & \multicolumn{2}{c}{Velocity} & \multicolumn{1}{c}{$B_{\ell}$} & \multicolumn{1}{c}{$N_{\ell}$}  \\
\multicolumn{1}{c}{no.} & \multicolumn{1}{c}{no.} & \multicolumn{1}{c}{Date}  & \multicolumn{1}{c}{(this work)} & \multicolumn{1}{c}{(Nidever)} & \multicolumn{1}{c}{} & \multicolumn{1}{c}{used} & \multicolumn{1}{c}{} & \multicolumn{1}{c}{} & \multicolumn{2}{c}{range} & \multicolumn{1}{c}{(\SI{}{\gauss})} & \multicolumn{1}{c}{(\SI{}{\gauss})}  \\
\multicolumn{1}{c}{} & \multicolumn{1}{c}{} & \multicolumn{1}{c}{}  & \multicolumn{1}{c}{(\SI{}{\kilo\meter\per\second})} & \multicolumn{1}{c}{(\SI{}{\kilo\meter\per\second})} & \multicolumn{1}{c}{} & \multicolumn{1}{c}{} & \multicolumn{1}{c}{} & \multicolumn{1}{c}{} & \multicolumn{2}{c}{(\SI{}{\kilo\meter\per\second})} & \multicolumn{1}{c}{} & \multicolumn{1}{c}{}  \\
\hline \\[-1.5ex]

1499 & 1 &  19oct10  & -10 & -10.166 & 23795 & 11140 &  N  &  6.059 $\times$ 10\sups{-01}  & -18 & -2 &  -0.6 $\pm$ 0.5  &  -0.6 $\pm$ 0.5 \\[3pt]
3093$^{+}$ & 1 &  10aug10  & -32.7 & -32.961 & 44842 & 13129 &  D (22/28) &  0.000 $\times$ 10\sups{+00}  & -40 & -25 &  -3.2 $\pm$ 0.2  &  -0.1 $\pm$ 0.2 \\[3pt]
7513 & 1 &  14nov06  & -28.3 & -28.674 & 36749 & 9872 &  N  &  7.340 $\times$ 10\sups{-01}  & -49 & -9 &  +2.5 $\pm$ 1.1  &  -0.3 $\pm$ 1.1 \\[3pt]
8159 & 1 &  27jan11  & -45.9 & -46.022 & 31246 & 10271 &  N  &  9.024 $\times$ 10\sups{-01}  & -54 & -38 &  -0.4 $\pm$ 0.3  &  +0.1 $\pm$ 0.3 \\[3pt]
12048 & 1 &  18dec10  & -50.7 & -50.971 & 22279 & 10271 &  N  &  4.743 $\times$ 10\sups{-02}  & -58 & -45 &  +1.0 $\pm$ 0.4  &  +0.4 $\pm$ 0.4 \\[3pt]
16537$^{+}$ & 1 &  01feb07  & 16.5 & 16.332 & 39693 & 11754 &  D (52/58) &  0.000 $\times$ 10\sups{+00}  & 11 & 23 &  -10.9 $\pm$ 0.2  &  -0.2 $\pm$ 0.2 \\[3pt]
 &  &  &  &  &  &  & M (3/58) &  &  &  &  & \\[3pt]
53721 & 1 &  31dec09  & 11.5 & 11.235 & 37281 & 9006 &  N  &  4.988 $\times$ 10\sups{-01}  & 5 & 18 &  +0.6 $\pm$ 0.2  &  +0.2 $\pm$ 0.2 \\[3pt]
67275$^{+}$ & 1 &  13may13  & -16 &  -16.542 $\pm$ 0.340  & 75344 & 8271 &  D (3/8) &  6.090 $\times$ 10\sups{-08}  & -38 & 5 &  +3.2 $\pm$ 0.5  &  -0.1 $\pm$ 0.5 \\[3pt]
 &  &  &  &  &  &  & M (2/8) &  &  &  &  & \\[3pt]
96901 & 1 &  17jul11  & -27.7 & -27.871 & 35324 & 9669 &  N  &  5.381 $\times$ 10\sups{-01}  & -47 & -9 &  -1.7 $\pm$ 0.9  &  -0.4 $\pm$ 0.9 \\[3pt]
100970 & 1 &  21jun11  & -91.1 &  -91.582 $\pm$ 0.188\sups{NS}  & 36392 & 10466 &  N  &  4.175 $\times$ 10\sups{-01}  & -104 & -77 &  +1.7 $\pm$ 0.6  &  -0.6 $\pm$ 0.6 \\[3pt]
107350$^{+}$ & 1 &  21jun10  & -16.6 & -16.833 & 21322 & 8764 &  D (68/91) &  0.000 $\times$ 10\sups{+00}  & -29 & -5 &  +14.8 $\pm$ 0.9  &  +0.0 $\pm$ 0.9 \\[3pt]
 &  &  &  &  &  &  & M (7/91) &  &  &  &  & \\[3pt]
109378 & 1 &  26nov10  & -20.7 & -20.873 & 37867 & 12151 &  N  &   4.703 $\times$ 10\sups{-01}  & -32 & -9 &  -0.9 $\pm$ 0.5  &  +0.2 $\pm$ 0.5 \\[3pt]
113357$^{*}$ & 1 &  15dec10  & -33.1 & -33.225 & 20777 & 11111 &  N  &  7.351 $\times$ 10\sups{-01}  & -40 & -27 &  +0.6 $\pm$ 0.4  &  +0.1 $\pm$ 0.4 \\[3pt]
113421 & 1 &  20oct10  & -13.3 & -13.399 & 21447 & 11135 &  N  &  5.690 $\times$ 10\sups{-01}  & -31 & 4 &  -1.9 $\pm$ 1.2  &  -0.5 $\pm$ 1.2 \\[3pt]

\hline
\end{tabular}
\end{table*}

\begin{table*}
\scriptsize
\caption{Chromospheric activity of planet-hosting stars in the BCool sample \protect\citep[Tables~5 and 1]{Marsden2014}.  $B-V$ and $V$ values are from \textit{Hipparcos}.  Where \protect\citet{Wright2004} has calculated an S-index, this is shown in column 3.  Chromospheric ages, periods and $\log(P_{rot}/\tau)$ have been determined using the equations presented in \protect\citet{Wright2004}.  Sample standard deviations are used as an indication in of various errors.}
\label{tab:derivedValuesBC}
\begin{tabular}{lcccccccccccccc}
\hline
\multicolumn{1}{c}{HIP} & \multicolumn{1}{c}{$B-V$} & \multicolumn{1}{c}{$V$} & \multicolumn{1}{c}{S-index} & \multicolumn{1}{c}{S-index} & \multicolumn{1}{c}{log($R^{\prime}_{HK}$)} & \multicolumn{1}{c}{Chromospheric} & \multicolumn{1}{c}{Chromospheric} & \multicolumn{1}{c}{Ca$_{\mbox{IRT}}$-index} & \multicolumn{1}{c}{H${_{\alpha}}$-index} & \multicolumn{1}{c}{$\log(P_{rot}/\tau)$} \\
\multicolumn{1}{c}{no.} & \multicolumn{2}{c}{(Hipparcos)} & \multicolumn{1}{c}{Wright.}  & \multicolumn{1}{c}{(this work)} & \multicolumn{1}{c}{} & \multicolumn{1}{c}{Age (\SI{}{\giga\year})} & \multicolumn{1}{c}{period (\SI{}{\day})} & \multicolumn{1}{c}{} & \multicolumn{1}{c}{} & \multicolumn{1}{c}{} & \multicolumn{1}{c}{} \\
\hline \\[-1.5ex]
1499 & 0.674 & 6.47 & 0.156 &  0.1567 $\pm$ 0.0014  &  -5.03$^{+0.01}_{-0.01}$  &  6.214$^{+0.221}_{-0.215}$  &  29.0$^{+0.3}_{-0.3}$  &  0.7149 $\pm$ 0.0016  &  0.2908 $\pm$ 0.0001  &  +0.336$^{+0.004}_{-0.004}$ \\[3pt]
3093 & 0.850 & 5.88 & 0.169 &  0.1724 $\pm$ 0.0007  &  -5.00$^{+0.00}_{-0.01}$  &  5.589$^{+0.203}_{-0.000}$  &  44.0$^{+0.4}_{-0.0}$  &  0.7065 $\pm$ 0.0007  &  0.3122 $\pm$ 0.0002  &  +0.324$^{+0.004}_{-0.000}$ \\[3pt]
7513 & 0.536 & 4.10 & 0.146 &  0.1565 $\pm$ 0.0009  &  -4.98$^{+0.01}_{-0.00}$  &  5.201$^{+0.000}_{-0.185}$  &  11.9$^{+0.0}_{-0.1}$  &  0.7498 $\pm$ 0.0014  &  0.2768 $\pm$ 0.0001  &  +0.316$^{+0.000}_{-0.004}$ \\[3pt]
8159 & 0.720 & 6.27 & 0.149 &  0.1445 $\pm$ 0.0013  &  -5.11$^{+0.01}_{-0.01}$  &  8.151$^{+0.269}_{-0.263}$  &  36.9$^{+0.3}_{-0.3}$  &  0.6900 $\pm$ 0.0012  &  0.2889 $\pm$ 0.0002  &  +0.366$^{+0.004}_{-0.004}$ \\[3pt]
12048 & 0.670 & 6.83 & 0.145 &  0.1461 $\pm$ 0.0019  &  -5.10$^{+0.01}_{-0.01}$  &  7.888$^{+0.263}_{-0.257}$  &  30.3$^{+0.3}_{-0.3}$  &  0.7067 $\pm$ 0.0013  &  0.2860 $\pm$ 0.0003  &  +0.362$^{+0.004}_{-0.004}$ \\[3pt]
16537 & 0.881 & 3.72 & 0.447 &  0.5357 $\pm$ 0.0019  &  -4.42$^{+0.00}_{-0.00}$  &  0.489$^{+0.000}_{-0.000}$  &  11.8$^{+0.0}_{-0.0}$  &  0.8566 $\pm$ 0.0017  &  0.3446 $\pm$ 0.0004  &  -0.262$^{+0.000}_{-0.000}$ \\[3pt]
53721 & 0.624 & 5.03 & 0.154 &  0.1501 $\pm$ 0.0014  &  -5.05$^{+0.01}_{-0.01}$  &  6.662$^{+0.233}_{-0.227}$  &  23.1$^{+0.2}_{-0.2}$  &  0.7519 $\pm$ 0.0014  &  0.2838 $\pm$ 0.0001  &  +0.343$^{+0.004}_{-0.004}$ \\[3pt]
67275 & 0.508 & 4.50 & 0.202 &  0.1760 $\pm$ 0.0001  &  -4.86$^{+0.00}_{-0.00}$  &  3.347$^{+0.000}_{-0.000}$  &  8.2$^{+0.0}_{-0.0}$  &  0.7431 $\pm$ 0.0003  &  0.2780 $\pm$ 0.0001  &  +0.259$^{+0.000}_{-0.000}$ \\[3pt]
96901 & 0.661 & 6.25 & 0.148 &  0.1537 $\pm$ 0.0005  &  -5.04$^{+0.00}_{-0.00}$  &  6.435$^{+0.000}_{-0.000}$  &  27.6$^{+0.0}_{-0.0}$  &  0.7476 $\pm$ 0.0009  &  0.2897 $\pm$ 0.0001  &  +0.340$^{+0.000}_{-0.000}$ \\[3pt]
100970 & 0.662 & 6.87 & 0.147 &  0.1521 $\pm$ 0.0041  &  -5.05$^{+0.03}_{-0.03}$  &  6.662$^{+0.717}_{-0.662}$  &  28.0$^{+0.7}_{-0.7}$  &  0.7154 $\pm$ 0.0007  &  0.2870 $\pm$ 0.0003  &  +0.343$^{+0.011}_{-0.012}$ \\[3pt]
107350 & 0.587 & 5.96 &  -  &  0.3330 $\pm$ 0.0014  &  -4.42$^{+0.00}_{-0.00}$  &  0.489$^{+0.000}_{-0.000}$  &  4.6$^{+0.0}_{-0.0}$  &  0.9147 $\pm$ 0.0005  &  0.3153 $\pm$ 0.0003  &  -0.262$^{+0.000}_{-0.000}$ \\[3pt]
109378 & 0.773 & 6.54 & 0.155 &  0.1534 $\pm$ 0.0007  &  -5.07$^{+0.01}_{-0.00}$  &  7.134$^{+0.000}_{-0.239}$  &  41.2$^{+0.0}_{-0.4}$  &  0.7024 $\pm$ 0.0012  &  0.3000 $\pm$ 0.0002  &  +0.351$^{+0.000}_{-0.004}$ \\[3pt]
113357 & 0.666 & 5.45 & 0.148 &  0.1528 $\pm$ 0.0021  &  -5.06$^{+0.01}_{-0.02}$  &  6.895$^{+0.484}_{-0.233}$  &  28.7$^{+0.5}_{-0.3}$  &  0.7154 $\pm$ 0.0014  &  0.2900 $\pm$ 0.0002  &  +0.347$^{+0.008}_{-0.004}$ \\[3pt]
113421 & 0.744 & 6.17 & 0.15 &  0.1494 $\pm$ 0.0029  &  -5.09$^{+0.02}_{-0.02}$  &  7.630$^{+0.521}_{-0.496}$  &  39.0$^{+0.7}_{-0.7}$  &  0.6861 $\pm$ 0.0006  &  0.2979 $\pm$ 0.0003  &  +0.359$^{+0.008}_{-0.008}$ \\[3pt]

\hline

\end{tabular}
\end{table*}

\label{lastpage}

\end{document}